\magnification=\magstep1
\hsize = 33pc
\vsize = 46pc

\def\bs{\baselineskip=14pt}
\bs

\tolerance 8000
\parskip=5pt

\font\grand=cmbx10 at 14.4truept
\font\grand=cmbx10 at 14.4truept\bigskip

\font\ninerm=cmr9

\def\q#1{[#1]}              
\def\bibitem#1{\parindent=8mm\item{\hbox to 17 mm{\q{#1}\hfill}}}

\def\BFP{{BFP}}
\def\DLNT{{DLNT}}
\def\DS{{DS}}
\def\EFS{{EFS}}
\def\F{{F}}
\def\FH{{FH}}
\def\FO{{FO}}
\def\FORTW{{FORTW}}
\def\GS{{GS}}
\def\He{{H}}
\def\KT{{KT}}
\def\KY{{KY}}
\def\OP{{OP}}
\def\P{{P}}
\def\PP{{PP}}
\def\R{{R}}
\def\first{{RSTS1}}
\def\second{{RSTS2}}
\def\STS{{STS}}
\def\TF{{TF}}
\def\War{{W}}
\font\extra=cmss10 scaled \magstep0
\setbox1 = \hbox{{{\extra R}}}
\setbox2 = \hbox{{{\extra I}}}
\setbox3 = \hbox{{{\extra C}}}
\def\RRR{{{\extra R}}\hskip-\wd1\hskip2.0 true pt{{\extra I}}\hskip-\wd2
\hskip-2.0 true pt\hskip\wd1}
\def\Real{\hbox{{\extra\RRR}}}    
\def\C{{{\extra C}}\hskip-\wd3\hskip2.5 true pt{{\extra I}}\hskip-\wd2
\hskip-2.5 true pt\hskip\wd3}
\setbox4=\hbox{{{\extra Z}}}


\def\G{{\cal G}}
\def\A{{\cal A}}
\def\B{{\cal B}}
\def\O{{\cal O}}
\def\M{{\cal M}}
\def\H{{\cal H}}
\def\N{{\cal N}}
\def\C{{\cal C}}
\def\tr{{\rm tr}}
\def\pa{{\partial}}

\pageno=0
\def\folio{
\ifnum\pageno<1 \footline{\hfil} \else\number\pageno \fi}

\baselineskip=12pt
\rightline{INS-Rep.-1123\break}
\rightline{ November 1995\break}
\rightline{ hep-th/9511118\break}
\vskip 1.0truecm

\bs

\vskip 1.0truecm

\centerline{\grand Regularization of  Toda lattices by Hamiltonian 
reduction}

\vskip 1.5truecm

\centerline{L\'aszl\'o Feh\'er \ and \  Izumi Tsutsui}

\vskip 0.8truecm

\centerline{\it Institute for Nuclear Study, University of Tokyo,}
\centerline{\it Midori-cho, Tanashi-shi, Tokyo 188, Japan}

\vskip 1.5truecm

\centerline{\bf Abstract}
\medskip

{\parindent=25pt
\baselineskip=13pt
\narrower\smallskip\noindent
\ninerm
The Toda lattice defined by the Hamiltonian
$H={1\over 2} \sum_{i=1}^n p_i^2 + \sum_{i=1}^{n-1} \nu_i e^{q_i-q_{i+1}}$
with  $\nu_i\in \{ \pm 1\}$, which  exhibits singular
(blowing up) solutions if some of the
$\nu_i=-1$,   can be viewed as the reduced system
following from a symmetry reduction of a  subsystem of the free particle
moving on the group $G=SL(n,\Real )$. The subsystem is $T^*G_e$, where
$G_e=N_+ A N_-$ consists of the determinant one
matrices with positive  principal minors, and the reduction is
based on the maximal nilpotent group
$N_+ \times N_-$.
Using the Bruhat decomposition we show that the full reduced system
obtained from  $T^*G$, which is perfectly regular,
contains $2^{n-1}$ Toda lattices.
More precisely, if $n$ is odd the reduced system contains
all the possible Toda
lattices having different signs for the $\nu_i$.
If $n$ is even, there exist two non-isomorphic reduced systems
with different constituent Toda lattices.
The Toda lattices occupy  non-intersecting open submanifolds
in the reduced phase space, wherein they
 are regularized by being glued together.
We find a model of the reduced phase space  as a hypersurface
in ${\Real}^{2n-1}$. If $\nu_i=1$ for all $i$, we  prove for $n=2,3,4$ that
the  Toda phase space  associated with  $T^*G_e$ is a
connected component of this hypersurface.
The generalization of the construction for  the other simple Lie
groups  is also presented.
}

\vfill\eject

\bs

\centerline{\bf 1.~Introduction}
\medskip

The Toda lattice and its generalizations
have been the subject of intense studies during the last
three decades and are widely recognized as one of
the  most important families of models in the theory of
integrable systems.
Various Toda  systems still attract attention today from viewpoints
spanning from differential geometry to conformal field theory.
The simplest of these systems is
the open (non-periodic) finite Toda lattice,
whose dynamics is generated by the Hamiltonian
$$
H(q,p)={1\over 2} \sum_{i=1}^n p_i^2 +\sum_{i=1}^{n-1} \nu_i e^{q_i-q_{i+1}}
\qquad(\nu_i:\,\hbox{non-zero constant})
\eqno(1.1)$$
on the phase space  $\Real^{2n}$ with  the canonical Poisson brackets.
This system, more precisely its center of mass reduction
defined by setting
$\sum_{i=1}^n p_i =\sum_{i=1}^n q_i =0$
and appropriately modifying  the Poisson bracket,
was historically the main source of the
Adler-Kostant-Symes (AKS) theory of integrable Hamiltonian systems.
In this Lie algebraic theory, whose most general form is based
on the classical r-matrix, the phase spaces of integrable systems
are realized as coadjoint orbits.
See e.g.~the  reviews in \q{\P,\second} and references therein.

In this paper we study an aspect of the Toda lattices defined in (1.1),
namely, their regularization in the singular case for which
some of the real constants $\nu_i$ are negative.
It is well-known that the Hamiltonian vector field of the open
Toda lattice is complete if all the constants $\nu_i$  are positive.
In contrast,   the Hamiltonian vector field
is  incomplete if $\nu_i<0$ for some $i$.
Indeed, if some $\nu_i<0$ then
it is intuitively clear that there should exist motions reaching
infinity at finite time due to the rapidly decreasing exponential
potential.
Surprisingly, as far as we know, a proof confirming  this
expectation only appeared very recently \q{\GS,\KY}.

Regularizing a singular, incomplete dynamical system means  embedding
it into a larger regular system whereby the incomplete trajectories get
smoothly continued from $-\infty$ to $+\infty$ in time.
Perhaps the most famous example is the regularization of the Kepler problem
obtained  by mapping the trajectories of  negative energy
into geodesics on a three-sphere $S^3$.
The regularization of an incomplete Toda lattice
will be achieved by realizing the system   as a restriction of a
larger Hamiltonian system having complete Hamiltonian
vector field  to an open submanifold of the phase space.
The possibility of such a regularization was emphasized by
Reyman and Semenov-Tian-Shansky \q{\first,\R,\second}.
Specifically, the  idea is to apply  Hamiltonian symmetry reduction
to the system describing a free particle moving on a Lie
group $G$, in our case $G=SL(n,\Real)$, in such a way
that the reduced system is  complete
and contains the Toda lattice on an open submanifold.
The way to define the symmetry reduction  emerges naturally  {}from
the orbital AKS interpretation of the Toda lattice \q{\first,\R,\second}.
Despite the idea being known for quite a while,
the completion of Toda lattices resulting from Hamiltonian reduction
has not yet been studied in detail.
In this paper it will be shown  that the reduced phase space
contains a {\it dense} open submanifold consisting of $2^{n-1}$ Toda 
lattices
which are in general singular on their own but have their
incomplete trajectories glued together ``at infinity'' represented
by the complement of this submanifold.
The full reduced system has an intricate structure,
which we will try to explore.

Incidentally, our  motivation originally comes from
our earlier studies \q{\FORTW} of the field theoretic version  of the
open Toda lattices  in which the method of Hamiltonian reduction was
used to derive them from the Wess-Zumino-Novikov-Witten model,
which is a field theoretic generalization of a free particle on a group.
In those investigations  we realized that the reduced phase space
automatically incorporates the singular solutions of the Toda field 
equations,
which have received much  attention
in the simplest $SL(2,\Real)$ case of the Liouville equation \q{\PP}.
Our present study of Toda lattices
with  finitely many degrees of freedom may serve as a first step
for field theoretic investigations in the same spirit.
In fact, some remarks on this  appeared already  in our
preliminary note in  \q{\TF} and a more extensive treatment in the
Liouville case is in preparation \q{\BFP}.

\noindent
The content  of this paper is as follows:

In Section 2 the orbital interpretation of the open Toda lattice
is presented to fix the background for later sections.
Two interpretations will be described
in the AKS formalism in correspondence with two different splittings
of the Lie algebra $sl(n,\Real)$.
The first is the lower triangular -- strictly upper triangular splitting
which is relevant also in the ``good sign'' case, that is,
the case of the Toda lattice with  $\nu_i>0$ for all $i$.
The second is a lower triangular -- pseudo-orthogonal splitting
generalizing the lower triangular -- orthogonal splitting used in the
standard case.
This second interpretation will allow us to make  contact with the
recent work in \q{\KY}.
The reader who is interested only in the regularization may skip this 
section.

Section 3 contains the symmetry reduction of the  system
on $T^* SL(n,\Real)$ that yields  the regularization of
the singular Toda lattices.
Our definition of the reduction is slightly different from
the one  proposed in \q{\first,\R,\second} since  the
strictly lower triangular $\times$ strictly upper triangular
symmetry algebra  is used instead of the
lower triangular $\times$ strictly upper triangular one,
but this leads to the same reduced phase space.
Then the  {\it Bruhat decomposition} of $SL(n,\Real)$,
whose relevance to  Toda lattices is well-known \q{\R,\FH},
is  applied to  describe  the content of the reduced system.
The reduced system turns out to contain $2^{n-1}$ Toda
lattices\footnote{${}^{1}$}{\ninerm
In particular,
a single   Toda lattice
does not form  a dense submanifold in the reduced phase space, which seems
to contradict  with  a claim in \q{\R}.}
as subsystems with various signs in the Hamiltonian in correspondence with
the non-intersecting open submanifolds of $SL(n,\Real)$ having
non-vanishing principal minors with fixed signs.
(For the precise statement, see the Theorem of Section 3.)
These Toda lattices
are glued together along lower dimensional
submanifolds
in the phase space which are related to the submanifolds  of $SL(n,\Real)$
with  some vanishing principal minors.

Our aim in Sections 4 -- 7 is to gain
a better understanding of the reduced system.
First we address the question of how many non-isomorphic possibilities are
permitted by the reduction.
There is a large freedom in choosing the value of the momentum map
(which corresponds to a  pair of matrices $I_-$, $I_+$ of the form in (2.2))
to fix  the constraints, but there are also equivalences induced by the
action of a certain  group of diagonal matrices.
In fact,
we find that if $n$ is odd then {\it all} reduced systems that arise are
isomorphic Hamiltonian systems, and if $n$ is even then there exist just
{\it two} non-isomorphic reduced systems.
This result is given by Proposition 1 in Section 4, and
Proposition 2  describes  the list of the Toda lattices
contained in the non-isomorphic reduced systems.

In Section 5 an involutive  symmetry of the reduced system,
which derives from  the outer automorphism of  $sl(n,\Real)$,
is exhibited.
With respect to this symmetry, we  establish
the behaviour of the  gauge invariant
functions defined  by the principal minors of $g\in SL(n,\Real)$.
The result, described in  Propositions 3 and 4,
proves to be useful when we  analyze  the reduced system.

Section 6 is devoted to deriving a model of the reduced phase space 
manifold,
which is of dimension $2(n-1)$, in the form of a hypersurface in
$\Real^{2n-1}$.
This is achieved with the aid of  a  global cross
section of the gauge orbits in the constrained manifold of the reduction.
We use an analogue of  the ``Drinfeld-Sokolov gauges''
familiar in the context of $n$-KdV systems \q{\DS}.
As a result, the hypersurface
is given by a polynomial equation.
This model should be useful  for further
investigating the {\it topology} of the reduced phase space,
although so far we have been able to carry  this out only
for some simple examples.

The examples alluded to in the above are contained in Section 7
and in Appendix A.
In Section 7 the $SL(2,\Real)$ and $SL(3,\Real)$ cases
are presented in detail.
In particular, in these cases the Toda lattice
with good sign  turns out to occupy a {\it connected
component} in the full reduced phase space.
It is appealing to conjecture that this is the case in general,
since in the good sign case regularization is not needed as
the Hamiltonian vector field is complete.
This conjecture is verified  in Appendix A  for $SL(4,\Real)$, too.

In Section 8  a short discussion of  the results
and an outlook are offered.
Finally,  the generalization of the
main construction of Section 3
is outlined for an arbitrary (real, split)
simple Lie algebra in Appendix B.


\bigskip
\medskip
\centerline{\bf 2.~Two orbital interpretations of the Toda lattice}
\medskip

We here explain two alternative orbital interpretations
of the Toda lattice studied in this paper.
For the Lie algebra $\G:=sl(n,\Real )$,
let us consider the decomposition
$$
\G=\G_{<0} + \G_0 +\G_{>0}
\eqno(2.1)$$
defined by the subalgebras of
strictly lower triangular, diagonal, and strictly upper triangular
traceless matrices, respectively.
Fix some elements
$$
I_+=\sum_{i=1}^{n-1} \nu_i^+ e_{i,i+1},
\qquad
I_-=\sum_{i=1}^{n-1} \nu_i^- e_{i+1,i},
\eqno(2.2)$$
where $e_{i,j}$ denotes  the $n\times n$ elementary matrix having
the entries $\left(e_{i,j}\right)_{k,l}=\delta_{i,k}\delta_{j,l}$
and  $\nu_i^\pm\neq 0$ ($i=1,\ldots, n-1$) are some arbitrarily
chosen real constants.
The phase space of the Toda lattice, which we denote by $M_e$, is defined
to be $M_e:=\G_0 \times \G_0$. The general element of $M_e$ is given
by a pair, $(q,p)$, of $n\times n$ diagonal, traceless matrices with
real entries,
$$
q={\rm diag}\left(q_1,\ldots,q_n\right),
\qquad
p={\rm diag}\left(p_1,\ldots,p_n\right),
\qquad
\sum_{i=1}^n q_i =\sum_{i=1}^n p_i=0.
\eqno(2.3)$$
The Toda dynamics is generated by the Hamiltonian $H_e$,
$$
H_e(q,p):={1\over 2}\tr\left(p^2\right)+\tr\left( I_- e^q I_+ e^{-q}\right)
={1\over 2}\sum_{i=1}^n p_i^2 + \sum_{i=1}^{n-1} \nu_i e^{q_i-q_{i+1}},
\qquad
\nu_i = \nu_i^- \nu_i^+,
\eqno(2.4)$$
by means of the symplectic structure $\omega_e=d \,\tr( p dq)$ on
$M_e$ leading to the Poisson brackets
$$
\{ q_i, q_k\}=\{ p_i, p_k\}=0,
\qquad
\{ q_i, p_k\} =\delta_{i,k} -{1\over n},
\quad
i,k=1,\ldots,n.
\eqno(2.5)$$
The corresponding equation of motion has the form
$$
{d q\over d t}=p,\qquad
{d p\over dt}=[ I_-\,,\, e^q I_+ e^{-q}].
\eqno(2.6)$$

The choice of $I_\pm$ in the above description of the system
has a certain
redundancy, since only the products $\nu_i =\nu_i^- \nu_i^+$ appear in
the Hamiltonian.  Moreover, using a constant shift of the coordinate  $q$,
we see that only the  signs of the $\nu_i$ matter.
Therefore, if desired,  we  could assume without loss of
generality that, say,
$$
\nu_i^-=1,
\qquad
\nu_i^+\in \{ \pm 1\}.
\eqno(2.7)$$

If some of the $\nu_i$ in (2.4)
are negative, then the Hamiltonian vector field
in (2.6) is {\it incomplete},
which means that there exist
trajectories blowing up to infinity at finite time.
An interesting  regularization
of this singularity will be
obtained from  Hamiltonian reduction.
Before discussing  this,  we wish to review  the interpretation
of the Toda lattice in  the  Adler-Kostant-Symes (AKS) framework
(see \q{\P,\second} and references therein).

In the  AKS approach to constructing integrable systems
one  starts  with some Lie algebra $\G$ (in our case $\G=sl(n,\Real )$)
and a splitting of $\G$ into a vector space direct sum of
two  Lie subalgebras $\A$ and $\B$,
$$\G=\A +\B.
\eqno(2.8)$$
The dual space $\G^*$ has the induced decomposition
$$
\G^* = \A^* + \B^*,\qquad
\A^*=\B^\perp,\quad  \B^*=\A^\perp,
\eqno(2.9)$$
where $\A^\perp\subset \G^*$ is the annihilator of $\A\subset \G$
with respect to the pairing
$\langle\ ,\ \rangle: \G^*\times \G\rightarrow \Real$.
The trick is to endow the space $\G^*$ with the Poisson bracket
$\{\ ,\ \}$ given by the direct difference of the Lie-Poisson brackets
on $\A^*$ and $\B^*$, that is,
for any smooth functions $f$, $h$ on $\G^*$,
$$
\{ f, h\}(\xi):=\langle \xi,  [ df_\B(\xi), dh_\B(\xi)]\rangle
-\langle \xi,  [ df_\A(\xi), dh_\A(\xi)]\rangle,
\qquad \forall \xi\in \G^*,
\eqno(2.10{\rm a})$$
where we have decomposed the differential  $df(\xi)\in \G$ of $f$ as
$$
df(\xi)=df_\A(\xi)  + df_\B(\xi),
\quad\hbox{with}\quad
df_\A(\xi)\in \A, \,\,\, df_\B(\xi)\in \B.
\eqno(2.10{\rm b})$$
The symplectic leaves of the Poisson manifold
$(\G^*, \{\ ,\ \})$  are the subspaces of the form
$$
\O = \O_\A + \O_\B,
\eqno(2.11)$$
where $\O_\A\subset \A^*$  (resp.~$\O_\B\subset \B^*$)
is a coadjoint orbit corresponding to the action of  $\A$ on
 $\A^*$ (resp.~$\B$ on $\B^*$).
Because of (2.10), here $\A^*$ (and thus also $\O_\A$)
is endowed with its own  Lie-Poisson bracket multiplied by $(-1)$.
The phase space $\O$ has a natural family of commuting Hamiltonians.
These  are given by  the restrictions of the
Casimir functions on $\G^*$, i.e., the elements of the ring
${\cal I}(\G^*)$ of functions invariant under the coadjoint action
of the Lie algebra $\G$ on $\G^*$.
The dynamics defined by $H\in {\cal I}(\G^*)$ has a generalized Lax form,
since the Hamiltonian vector field $\chi_H$ on $\G^*$ generated by
$H\in {\cal I}(\G^*)$
through the  Poisson bracket in (2.10)  is given by
$$
\chi_H(\xi) = -\left({\rm ad}^* dH_\A(\xi) \right) (\xi ) =
\left({\rm ad}^* dH_\B(\xi) \right) (\xi),
\qquad \forall \xi\in \G^*\,,
\eqno(2.12)$$
where  $({\rm ad}^* X)(\xi)$ stands for the coadjoint action of
$X\in \G$ on $\xi\in \G^*$.
The integration of  Hamilton's equation
$$
{d\over dt} \xi(t)=\chi_H(\xi(t))
\eqno(2.13)$$
can be reduced  to  a factorization problem in the connected
Lie  group $G$ associated with $\G$.
In fact \q{\P,\second},  if one defines 
$X_0:=dH(\xi_0)$ and considers the factorization
$$
e^{tX_0}=a(t)b(t),
\eqno(2.14)$$
where $a(t)$ (resp.~$b(t)$) belongs to the Lie subgroup
of $G$ corresponding   to the Lie subalgebra $\A\subset \G$
(resp.~$\B\subset \G$),
then one finds the  solution of (2.13) through the initial value
$\xi_0$ in the form
$$
\xi(t)= {\rm Ad}^*_{a^{-1}(t)}\,  \xi_0 ={\rm Ad}^*_{b(t)}\, \xi_0,
\eqno(2.15)$$
with ${\rm Ad}^*_g$ denoting the coadjoint action of $g\in G$ on $\G^*$.

In general the factorization problem (2.14)
has a solution only locally, signalling the possible incompleteness
of the Hamiltonian vector field, which is of course
tangent to any orbit $\O$.
Many interesting integrable systems can be described in the
above framework.
The  interpretation of the Toda lattice presented  in Subsection 2.1
will reappear in the subsequent Hamiltonian reduction treatment,
while the alternative interpretation given in
Subsection 2.2  allows  us to make contact with
the recent work in \q{\KY}.

\medskip
\noindent
{\bf 2.1.~The lower triangular  --- strictly upper triangular  splitting}
\medskip

Let  $\G:=sl(n,\Real)$ and  $G:=SL(n,\Real )$.
Consider the splitting (2.8) with $\A$ being the strictly upper
triangular nilpotent subalgebra,
and $\B$ the lower triangular Borel subalgebra,
$$
\A:=\G_{>0},
\qquad
\B:=\G_{\leq 0}=\G_0 + \G_{<0}.
\eqno(2.16)$$
Denote by $N_+$ and $B_-$ the connected subgroups of $G$ corresponding
to $\A$ and $\B$.
Identifying $\G^*$ with $\G$ using the scalar product provided
by ordinary matrix trace, we have
$$
\A^*=\G_{<0},
\qquad
\B^*=\G_{\geq 0}.
\eqno(2.17)$$
Let us  choose $\O_\A$ to be the
one point coadjoint orbit (character) of $N_+$ given by $\O_\A=\{ I_-\}$
with $I_-$  in (2.2), and
choose $\O_\B$ to be the coadjoint orbit of $B_-$ through
the point $I_+\in \B^*$.
The space  $\O=\O_\A +\O_\B$ can then be parametrized as
$$
\O=\left\{\, J_e=I_- +p+e^q I_+ e^{-q}\,\vert\,
\forall\, (q,p)\in \G_0\times \G_0\,\right\}.
\eqno(2.18)$$
We may identify $\O$ with the Toda phase space $M_e$ defined above.
The Hamiltonian $H_e$ in (2.4) becomes
$$
H_e(J_e) ={1\over 2} \tr\left(J_e^2\right).
\eqno(2.19)$$
A commuting family of
independent Hamiltonians on $\O$ is provided by the set
$H_i(J_e)={1\over i} \tr\left(J_e^i\right)$ for $i=2,\ldots, n$.
This implies the integrability of the Toda equation (2.6),
which is  generated by  $H_e=H_2$ and can be re-casted according to
(2.12) in the Lax form
$$
{d J_e\over dt}= [J_e\,,\, (J_e)_{>0}].
\eqno(2.20)$$
According to (2.14), (2.15),
the integration algorithm 
corresponding to this orbital interpretation
of the Toda lattice
uses the Gauss decomposition
$$
e^{tJ_e}=n_+(t) b_-(t),
\qquad
n_+(t)\in N_+,\quad b_-(t)\in B_-,
\eqno(2.21)$$
which is valid if $e^{tJ_e}$ belongs to 
a neighbourhood of the identity on the group $SL(n,\Real)$ (see Section 3).

\medskip
\noindent
{\bf 2.2.~The lower triangular --- pseudo-orthogonal splitting}
\medskip

In the ``good sign'' case for which $\nu_i>0$ the Toda vector
field is known to be  complete.
This can be seen with the aid of the alternative
orbital interpretation \q{\P,\second} given in terms of the
Iwasawa decomposition of $\G=sl(n,\Real )$,
$$
\G = o(n,\Real ) +\G_{\leq 0},
\eqno(2.22)$$
which corresponds to a global decomposition of $G$ \q{\He,\War}.
We below show that replacing $\nu_i>0$ in the Toda Hamiltonian
by $\nu_i\neq 0$ having arbitrary signs amounts to  replacing
$o(n,\Real)$ in the splitting (2.22) by a  pseudo-orthogonal
Lie algebra. The singularity of the resulting  Toda lattices
is related to the fact that the pseudo-orthogonal
analogue of the Iwasawa decomposition is
not a global decomposition.

Given $I_\pm$ in (2.2), we can find a diagonal matrix
$S={\rm diag}\left(s_1, s_2,\ldots, s_n\right)$
for which
$$
I_+ = S^{-1} (I_-)^T S.
\eqno(2.23)$$
We use the notation  $X^T$
to denote the transpose of any matrix $X$.
In fact, $S$ in (2.23)  is unique up to an overall constant,
which may be fixed, e.g., by setting $s_1=1$.
After choosing $S$, we define $o_S$ to be the subalgebra
of $sl(n,\Real )$ preserving the symmetric form on $\Real^n$
associated with  $S$,
$$
o_S=\left\{\, X\in sl(n,\Real )\,\vert\, X^T=-S X S^{-1}\,\right\}.
\eqno(2.24)$$
Up to isomorphism,  the Lie algebra $o_S$
depends only on the signature of the symmetric form associated with $S$.
As is readily seen, a splitting of $\G$ is defined by
$$
\A:= o_S,
\quad
\B:= \G_{\leq 0}.
\eqno(2.25)$$
The space $\B^*$ in (2.9) is  now identified as
$$
\B^*=o_S^\perp = \left\{\, L\in \G\,\vert\, L^T=SLS^{-1}\,\right\}.
\eqno(2.26)$$
To re-obtain the Toda lattice, we now consider
the coadjoint orbit $\tilde \O$ of the group $B_-$ through the
element $I:=(I_- + I_+)\in \B^*$.
This  orbit can be parametrized as
$$
\tilde \O=\left\{\, L=e^{q/2} I_+ e^{-q/2} + p + e^{-q/2} I_- e^{q/2}\,
\vert \, \forall (q,p) \in \G_0\times \G_0\,\right\}.
\eqno(2.27)$$
To compare with the space $\O$ in (2.18),
we notice  that the mapping
$$
\phi: \O \rightarrow \tilde \O,
\qquad
\phi: J_e=(I_- + p + e^q I_+ e^{-q}) \mapsto L=e^{-q/2} J_e e^{q/2}
\eqno(2.28)$$
is a symplectomorphism  if $\O$ is endowed with the Lie-Poisson bracket
of $\B^*$ and $\tilde \O$ is by definition
endowed with $1\over 2$-times the Lie-Poisson
bracket of $\B^*$.
In the $\tilde \O$ realization the commuting Hamiltonians are given
by the trace of the powers of the Lax matrix $L$.

{}From  (2.12) with the present conventions,
the flow generated on $o_S^\perp$
by the Hamiltonian $H_k(L)={1\over k} \tr\left(L^k\right)$
takes the form
$$
{d L\over dt}= {1\over 2} \bigl[L\,,\, \left(L^{k-1}\right)_{o_S}\bigr]
\eqno(2.29{\rm a})$$
with
$$
\left(L^{k-1}\right)_{o_S}=
 \left(L^{k-1}\right)_{>0}-
S^{-1}\left(\left(L^{k-1}\right)_{>0}\right)^T S
=\left(L^{k-1}\right)_{>0} - \left(L^{k-1}\right)_{<0}.
\eqno(2.29{\rm b})$$
The left hand side of this equation is defined using the
splitting $\G=o_S +\G_{\leq 0}$ while  the right hand side
is defined using the decomposition of $\G$ in (2.1).
One may check that
the Toda equation (2.6) is recovered from (2.29) with $k=2$  under
restriction to the orbit $\tilde \O$.

The hierarchy on $o_S^\perp$
given in (2.29) is actually the same as the
``Toda hierarchy with indefinite metric''
introduced  in \q{\KY}  generalizing
the ``full symmetric  Toda hierarchy'' studied in \q{\DLNT},
for which  $S$ is  the identity matrix.
This follows by noting that any Lax matrix  $L\in o_S^\perp$
can be written in the form  $L=l S$ with  a symmetric matrix
$l$, and this is precisely the form of the Lax matrix postulated in \q{\KY}.
In this reference  the explicit solution of equation
(2.29) with $k=2$  was obtained
in terms of the solution of a factorization problem involving
the pseudo-orthogonal group whose Lie algebra is $o_S$.
The above presentation permits us to view this factorization
as a special case of the AKS scheme described earlier.
More importantly, the results of  \q{\KY} (see also \q{\GS})
show that if some $\nu_i<0$ in (2.4) then the Toda
system  indeed has trajectories that blow up to infinity at finite time.
To see this we recall that if $L_0\in \tilde\O$ has a
complex eigenvalue
then the solution of the Toda equation
$$
{d L\over dt}={1\over 2}\left[L\,,\, L_{>0}-L_{<0}\right]
\eqno(2.30)$$
through the initial value $L_0$ blows up to infinity (in $q$ space)
at finite time \q{\KY,\GS}.
Clearly,  if the energy $H_2={1\over 2} \tr(L^2)$ is negative
then $L$ must have a complex eigenvalue.
On the other hand, since ${1\over 2}\tr(L^2)=H_e(q,p)$, we see from
the formula of $H_e(q,p)$ in (2.4) that there exist initial data
with negative energy whenever $\nu_i<0$ for some $i$.
Therefore in such a case the Toda lattice is
singular\footnote{${}^{2}$}{\ninerm
The  existence of a complex eigenvalue of $L_0$
is sufficient but not necessary \q{\KY,\GS} for the singularity
of the solution.
A necessary and sufficient condition is given in \q{\GS}.}  (incomplete).
The  blowing up of the
solutions will be illustrated in examples later.

\medskip
\noindent
{\it Remark 1.}
We can define a commuting hierarchy on $\B^*$
by restricting the Casimir functions on $\G^*$
to $\B^*\equiv\A^\perp + \{ I\}$ using this identification
for any  choice of $\A$ in (2.8) and any  character
$I\in \B^\perp$ of $\A$.
The resulting hierarchies are not isomorphic in general as
illustrated by the  hierarchies
on $\B^*=\left(\G_{\leq 0}\right)^*$ studied in \q{\DLNT,\KY}.
Another  hierarchy on $\left(\G_{\leq 0}\right)^*$ is
the ``full Kostant-Toda lattice''  investigated  in \q{\EFS},
which is obtained  by choosing $\A=\G_{>0}$ and $I:=I_-$ in (2.2).
Of course the restrictions of these hierarchies
to some special  coadjoint orbits in $\B^*$ can be (and are) isomorphic
in some cases.

\bigskip
\medskip
\centerline{\bf 3.~Regularization by Hamiltonian reduction}
\medskip

In this  section we consider a Hamiltonian symmetry reduction of the system
describing a free particle on the group $G=SL(n,\Real )$.
This  leads to a reduced system that contains  the Toda lattice defined
in the preceding section. More precisely, it turns out that the reduced
system, which is perfectly regular,  contains not only one but
$2^{n-1}$ Toda lattices that are glued together in such a way to
provide a natural regularization of their singularities.
The idea  that Hamiltonian reduction can be used to regularize
singular, incomplete Toda  systems arising in the AKS framework
goes back to Reyman and Semenov-Tian-Shansky  \q{\first,\R}.
The details of this  regularization
has not been  however worked out previously.

A model of the  free particle on the group $G$
is furnished  by the  Hamiltonian system $(\M,\Omega,\H)$ as follows.
The phase space $\M$ is the cotangent bundle of the group,
$$
\M = T^*G \simeq \left\{(g,J)\, \vert\,\, g \in G, \, J 
\in {\cal G}\,\right\},
\eqno(3.1)
$$
where $\G^*$ is identified with $\G$ using the scalar product, and
the identification $T^*G\simeq G\times {\cal G}$ is defined
by right translations on $G$.
The fundamental Poisson brackets are
$$
\{g_{i,j}\, , g_{k,l}\} = 0, \quad
      \{g_{i,j}\, , \tr(T^aJ) \} = (T^a g)_{i,j}, \quad
\{\tr(T^aJ)\, , \tr (T^bJ)\} = \tr ([T^a, T^b] J),
\eqno(3.2)
$$
with $T^a$ being a basis of  ${\cal G}=sl(n,\Real )$.
These derive from the symplectic form $\Omega$,
$$
\Omega = d\, \tr \left( J dg\, g^{-1}\right).
\eqno(3.3)$$
The Hamiltonian $\H$ is taken to be
$$
\H = {1\over2} \tr\left( J^2\right),
\eqno(3.4)
$$
which yields the dynamics,
$$
{d g\over dt} = \{ g \, , \H\} = Jg, \qquad {d J\over dt} = 
\{ J\, , \H\} = 0.
\eqno(3.5)
$$
Hence the particle, whose position is given by
$g(t)\in G$,  moves according to
$${d\over{dt}}\left({d g\over dt} g^{-1}\right) = 0.
\eqno(3.6)$$

Note that $J$ is the infinitesimal generator, i.e., the momentum map,
for the action of $G$ on $\M$ defined by left translations, while the
action of $G$ defined by right translations is generated  by
the momentum map $\tilde J: \M\rightarrow \G$,
$$
\tilde J(g,J)=-g^{-1} Jg.
\eqno(3.7)$$

We wish to consider symmetry reduction using the subgroup
$$
N:=N_+\times N_-
\eqno(3.8{\rm a})$$ of the full  symmetry group,
where $N_+$ and $N_-$
are  the subgroups of $G$ associated with  the Lie algebras $\G_{>0}$ and
$\G_{<0}$, respectively.
The group $N$ acts on  $\M$ according to
$$
(n_+,n_-): (g,J)\mapsto (n_+ g n_-^{-1}, n_+Jn_+^{-1}),
\qquad \forall\, (n_+,n_-)\in N, \quad (g,J)\in \M.
\eqno(3.8{\rm b})$$
Identifying
the dual of the Lie algebra $\N=\G_{>0}\times \G_{<0}$ of $N$
as $\N^*={\cal G}_{<0}\times \G_{>0}$,
the momentum map $\Phi: \M\rightarrow \N^*$ corresponding to
the symmetry in (3.8) is given by
$$
\Phi(g,J)=\left( J_{<0}, \tilde J_{>0}\right),
\eqno(3.9)$$
where $J=J_{<0}+J_0+J_{>0}$ according to (2.1) and similarly for $\tilde J$.
We define the  symmetry reduction by fixing the value of the momentum map
$\Phi$ to be $(I_-, -I_+)\in \N^*$,
where $I_\pm$ are the matrices in (2.2).
The reduced phase space is obtained as the factor space
$$
\M^{\rm red}(I_-,I_+)=\M^{\rm c}(I_-,I_+)/N,
\eqno(3.10)$$
with $$
\M^{\rm c}(I_-,I_+):=\Phi^{-1}\left(I_-, -I_+\right).
\eqno(3.11{\rm a})$$
We shall often write simply $\M^{\rm c}$ and $\M^{\rm red}$
omitting the argument $(I_-,I_+)$.
In Dirac's terminology, the constraints defining $\M^{\rm c}\subset \M$,
$$
J_{<0}=I_-,\qquad \tilde J_{>0}=-I_+,
\eqno(3.11{\rm b})$$
are {\it first class} and thus in (3.10)  we have to factorize by
the gauge group $N$.
(In other words, $(I_-,-I_+)\in \N^*$ is a character.)
It is  known that $\M^{\rm red}$ is a smooth manifold since
$(I_-, -I_+)$ is a regular value of the momentum map and the action
of $N$ on $\M^{\rm c}$ is free and proper \q{\R}.
Our task is to describe the reduced Hamiltonian system
$(\M^{\rm red}, \H^{\rm red}, \Omega^{\rm red})(I_-,I_+)$, where
$\Omega^{\rm red}$ and $\H^{\rm red}$ on $\M^{\rm red}$ are
naturally inherited from $\Omega$ and $\H$ on $\M$.

Let ${\bf M}$ be the finite subgroup
of $G=SL(n,\Real )$  consisting of diagonal matrices with entries
$\pm 1$.
An element $m\in {\bf M}$ is a diagonal matrix
$$
m={\rm diag}\left(m_1,m_2,\ldots,m_{n}\right),
\quad
m_i=\pm 1,\quad \prod_{i=1}^n m_i =1,
\eqno(3.12)$$
and hence ${\bf M}$ has $2^{n-1}$ elements. For any $m\in {\bf M}$,
define $G_m\subset G$ by
$$
G_m := N_+ m A N_-,\qquad A:=\exp\left(\G_0\right),
\eqno(3.13)$$
where $A$ is the subgroup of $G$
consisting of diagonal  matrices with positive entries.
As a special case of the {\it Bruhat (Gelfand-Naimark) decomposition}
 of semisimple
Lie groups \q{\He,\War},  we have
$$
G=\cup_m G_m \cup G_{\rm low}\qquad \hbox{(disjoint union)},
\eqno(3.14)$$
where the ``big cell'' $\cup_m G_m$
is an open, dense submanifold in $G$
and $G_{\rm low}$ is the  union of
certain lower dimensional
submanifolds of $G$.
The decomposition $g=n_+ man_-$ of any $g\in G_m$, with
$n_\pm \in N_\pm$, $a\in A$,  is unique.
The open submanifold
$G_m\subset G$ is diffeomorphic to  $N_+\times A\times  N_-$.
The detailed structure of $G_{\rm low}$ will not be used in this paper.

The big cell $\cup_m G_m$ of $SL(n,\Real )$
is the open submanifold  of  determinant  one
matrices with  non-vanishing principal minors.
The element $m\in {\bf M}$
fixes the signs of the principal minors,
which is possible in $2^{n-1}$ different ways.
Correspondingly, $G_{\rm low}$
consists of the matrices with unit determinant and
at least one vanishing principal minor.

Consider now the decomposition of $\M=T^*G$  induced by
the Bruhat decomposition of $G$,
$$
\M=\cup_m \M_m \cup \M_{\rm low}\qquad \hbox{(disjoint union)}.
\eqno(3.15)$$
The cells  $\M_m=T^*G\vert_{G_m}$ and
$\M_{\rm low}=T^*G\vert_{G_{\rm low}}$ in
this partitioning of $\M$
are {\it invariant} submanifolds with respect to the action of the
symmetry group  $N_+\times N_-$.
Therefore the corresponding partitioning
$$
\M^{\rm c}=\cup_m \M^{\rm c}_m \cup \M^{\rm c}_{\rm low}
\eqno(3.16)$$
of $\M^{\rm c}$  induces a decomposition
of the reduced phase space,
$$
\M^{\rm red}=\cup_m \M^{\rm red}_m \cup \M^{\rm red}_{\rm low}
\qquad\hbox{(disjoint union)}.
\eqno(3.17)$$
It is clear
that $\M^{\rm red}_m$ is an open submanifold in $\M^{\rm red}$
for any $m\in {\bf M}$ and  $\M^{\rm red}_{\rm low}$ is
a union of lower dimensional submanifolds.

For $I_+$ and $m\in {\bf M}$ given, define
$$
I_+^m := mI_+ m^{-1}.
\eqno(3.18)$$
The following theorem asserts that the subsystem
$$
\left(\M^{\rm red}_m, \Omega^{\rm red}_m, \H^{\rm red}_m\right),
\qquad m\in {\bf M},
\eqno(3.19)$$
of the reduced Hamiltonian system
$\left(\M^{\rm red}, \Omega^{\rm red}, \H^{\rm red}\right)$
obtained by restriction to the submanifold
$\M_m^{\rm red}\subset \M^{\rm red}$ is
a Toda lattice of the type defined  in Section 2.

\smallskip\noindent{\bf Theorem.}
{\it The subsystem  of the reduced system  in (3.19)
is a Toda lattice, whose phase
space in the realization provided by the lower triangular ---
strictly upper triangular
splitting  is given by the space of Lax matrices $J_m$ of the form
$$
J_m= I_- + p + e^q I_+^m e^{-q}, \qquad \forall\, (q,p)\in \G_0\times \G_0.
\eqno(3.20)$$}
\noindent
{\it Proof.}
{}From (3.11),
the submanifold $\M^{\rm c}_m\subset \M^{\rm c}$ can be written as
$$
\M^{\rm c}_m=\left\{\,(g ,J)\,\vert\, g=n_+me^qn_-,\,\,
 n_\pm \in N_\pm,\,\, q\in \G_0,
\,\,\, J_{<0}=I_-,\,\, \left(g^{-1} J g\right)_{>0}=I_+\,\right\}.
\eqno(3.21)$$
Hence  a model of
$\M_m^{\rm red}=\M^{\rm c}_m/(N_+\times N_-)$ is given by  the local gauge
section $\C_m\subset \M^{\rm c}_m$,
$$
\C_m=\left\{\,(me^q ,J)\,\vert\,   q\in \G_0,
\quad J_{<0}=I_-,\,\, \left(e^{-q}m^{-1} J me^{q}\right)_{>0}=I_+\,\right\}.
\eqno(3.22)$$
Solving the condition in (3.22) for $J$
yields that $J=J_m$  with $J_m$ given in (3.20).
Since $\C_m$ is a model of $\M^{\rm red}_m$, and  since once $J_m$ is given
the values of both $q$ and $p$ are uniquely determined,
 we see that the manifold
$$
\O_m:=\left \{ \, J_m=I_- + p + e^q I_+^m e^{-q}\,\vert\, \forall\, (q,p)\in
\G_0\times \G_0\,\right\}
\eqno(3.23)$$
is an equally good model of the manifold $\M_m^{\rm red}$.
We can regard $\O_m$ as a Toda orbit of the type in (2.18), where we simply
replace $I_+$ by $I_+^m$.
Upon the identification $\O_m\simeq\M^{\rm red}_m$,
the symplectic structure on
the coadjoint orbit $\O_m$ of $N_+\times B_-$ is
the same as  the symplectic structure $\Omega^{\rm red}_m$ 
of $\M^{\rm red}_m$
following from  the Hamiltonian reduction.
In fact,
$$
\Omega^{\rm red}_m=d\, \tr\left(p dq\right).
\eqno(3.24)$$
The commuting Hamiltonians provided by the AKS scheme  coincide
with those induced by Hamiltonian reduction from the  commuting
Hamiltonians
$$
\H_k (J) ={1\over k}\tr\left(J^k\right),\qquad k=2,\ldots, n,
\eqno(3.25)$$
on $\M$.
The proof of the theorem is completed by noting that
the Toda Hamiltonian
$$
\H_m^{\rm red}(J_m)={1\over 2}\tr\left(J_m^2\right)=
{1\over 2} \sum_{i=1}^n p_i^2 +
\sum_{i=1}^{n-1} \mu_i e^{q_i-q_{i+1}},
\eqno(3.26{\rm a})$$
where
$$
\mu_i=m_i m_{i+1}\nu_i,
\qquad \nu_i=\nu_i^+\nu_i^-,
\eqno(3.26{\rm b})$$
arises from (3.25) for $k=2$. {\it Q.E.D.}

The Toda flow on the manifold
$\O_m\simeq \M^{\rm red}_m$
is governed by the equation
$$
{d q\over dt}=p,
\qquad
{d p\over d t}=[I_-,e^q I_+^m e^{-q}].
\eqno(3.27)$$
This flow on $\O_m$ is incomplete (singular)
if and only if the value of
 $\mu_i$ is negative for some $i$.
But there is no singularity in the full reduced system.
The incompleteness of the system
on the coadjoint orbit $\O_m=\M_m^{\rm red}$
is a manifestation of the fact that
the particle may leave the submanifold
$\M^{\rm red}_m\subset \M^{\rm red}$ at finite time.
In concrete terms,
this simply means that the trajectory
of the free particle on $G$ determined by  an initial value
$(me^q,J_m)$ at $t=0$, which is
given explicitly  by
$$
g(t) = e^{t J_m} m e^q,
\eqno(3.28)$$
may leave the open submanifold $G_m$.
(Here we used the fact that the flow of the reduced system
is obtained by just projecting the original  flow on
$\M^{\rm c}\subset \M$ to $\M^{\rm red}$.)
This happens at time $t_*$ if  a principal minor of $g(t_*)$ vanishes,
which corresponds to $q$ reaching infinity at $t=t_*$ from the perspective
 of $G_m\subset G$.
In conclusion,
the embedding of $\O_m\simeq \M_m^{\rm red}$ into $\M^{\rm red}$
for all $m\in {\bf M}$ provides
a regularization of the singular Toda lattices whereby their
blowing up trajectories  are glued together ``at infinity''.

\bigskip
\medskip
\centerline{\bf 4.~The set of non-isomorphic reduced systems}
\medskip

We have associated the reduced Hamiltonian
system $(\M^{\rm red}, \Omega^{\rm red}, \H^{\rm red})$ with any
fixed pair $(I_-, I_+)$ of matrices given in (2.2).
The set of reduced systems so obtained is apparently parametrized
by the different choices of $(I_-,I_+)$, but this parametrization
is highly redundant since  different choices may lead to isomorphic
reduced systems.
The aim of this section is to describe the non-isomorphic
reduced systems obtained from $(\M,\Omega,\H)$,
$\M=T^* SL(n,\Real )$.
The situation turns out to be different depending on
whether  $n$ is odd or even.
For $n$ odd, we find that {\it all} reduced systems are isomorphic
and contain a copy of the Toda lattice  with Hamiltonian
$$
\H^{\rm red}_m(q,p):={1\over 2}\tr\left(p^2\right)+
\tr\left( I_- e^q I^m_+ e^{-q}\right)
={1\over 2}\sum_{i=1}^n p_i^2 + \sum_{i=1}^{n-1} \mu_i e^{q_i-q_{i+1}}
\eqno(4.1)$$
for any choice of ${\rm sign}( \mu_i)$ for $i=1,\ldots, (n-1)$.
For $n$ even, there exist  {\it two} non-isomorphic reduced systems
whose constituent Toda lattices  will be given by Proposition 2 later.

To  search for the non-isomorphic reduced systems, let us consider
 a pair, $(D, \bar D)$, of real, diagonal,
non-singular matrices of equal determinant,
$$
D={\rm diag}\left(d_1,\ldots, d_n\right),
\qquad
\bar D={\rm diag}\left(\bar d_1,\ldots, \bar d_n\right),
\qquad
\det D=\det \bar D.
\eqno(4.2)$$
Denote by $\Delta$ the group whose elements are such pairs,
$(D_1, \bar D_1) (D_2, \bar D_2)=(D_1D_2, \bar D_1 \bar D_2)$.
To any $(D,\bar D)\in \Delta$, define $\phi_{(D,\bar D)}$ to be
the {\it symmetry transformation} of the Hamiltonian system 
$(\M,\Omega,\H)$
given by
$$
\phi_{(D,\bar D)}: (g,J)\mapsto (Dg\bar D^{-1}, DJD^{-1})
\qquad \forall\, (g,J)\in \M.
\eqno(4.3)$$
Focusing  on the  constrained manifold  $\M^{\rm c}(I_-,I_+)\subset \M$,
we see that $\phi_{(D,\bar D)}$ in (4.3) induces an isomorphism
$$
\phi_{(D,\bar D)}: \M^{\rm c}(I_-,I_+)\rightarrow
\M^{\rm c}(I_-^D, I_+^{\bar D}),
\qquad
I_-^D=DI_-D^{-1},
\quad
I_+^{\bar D}=\bar D I_+ \bar D^{-1}.
\eqno(4.4)$$
Because of
$$
DN_+ D^{-1}=N_+,
\qquad
\bar DN_-\bar D^{-1}=N_-,
\eqno(4.5)$$
$\phi_{(D,\bar D)}$ in (4.4) maps gauge orbits to gauge orbits.
Hence it gives rise to a mapping
$$
\hat \phi_{(D,\bar D)}: \M^{\rm red}(I_-,I_+)\rightarrow
\M^{\rm red}(I_-^D, I_+^{\bar D}),
\eqno(4.6)$$
which is an isomorphism between the corresponding reduced
systems.

In order to find the non-isomorphic reduced systems,
we have to study the orbits of the group $\Delta$ acting on the
set of matrices $(I_-,I_+)$ as
$\Delta\ni (D,\bar D): (I_-, I_+)\mapsto (I_-^D, I_+^{\bar D})$.

\medskip
\noindent
{\bf Proposition 1.} {\it
If $n$ is odd, then the set of pairs $(I_-,I_+)$
of the form in (2.2) is a single orbit of the group $\Delta$.
If $n$ is even, then the set of pairs $(I_-,I_+)$
consists of two orbits of $\Delta$.}

\medskip
\noindent
{\it Proof.}
For arbitrarily fixed matrices,
$$
I_-=\sum_{i=1}^{n-1} \nu_i^- e_{i+1,i},
\quad
I_+=\sum_{i=1}^{n-1} \nu_i^+ e_{i,i+1},
\quad
\hat I_-=\sum_{i=1}^{n-1} \hat \nu_i^- e_{i+1,i},
\quad
\hat I_+=\sum_{i=1}^{n-1} \hat \nu_i^+ e_{i,i+1},
\eqno(4.7)$$
consider
$$
\hat I_-=DI_- D^{-1},
\qquad
\hat I_+ =\bar D I_+ \bar D^{-1}
\eqno(4.8)$$
as an equation for $(D,\bar D)\in \Delta$.
In detail, (4.8) requires
$$
\hat\nu_i^-=d_{i+1}\nu_i^- d_i^{-1}
\qquad\hbox{and}\qquad
\hat\nu_i^+=\bar d_{i}\nu_i^+ \bar d_{i+1}^{-1}
\qquad \forall\, i=1,\ldots, (n-1),
\eqno(4.9)$$
which leads to
$$
d_i=d_1 \prod_{l=1}^{i-1} {\hat \nu_l^-\over \nu_l^-},
\qquad
\bar d_i=\bar d_1 \prod_{l=1}^{i-1} { \nu_l^+\over \hat \nu_l^+}
\qquad
\forall\, i=2,\ldots, n.
\eqno(4.10)$$
Introducing the notation
$$
\nu^\pm :=\prod_{i=1}^{n-1} \left(\nu_i^\pm\right)^{n-i}\,,
\qquad
\hat \nu^\pm :=\prod_{i=1}^{n-1} \left(\hat \nu_i^\pm\right)^{n-i},
\eqno(4.11)$$
it follows that
$$
\det D=(d_1)^n { \hat \nu^-\over \nu^-}\,,
\qquad
\det \bar D= (\bar d_1)^n { \nu ^+ \over \hat \nu^+}\,.
\eqno(4.12)$$
If $n$ is odd, a solution of (4.9)  satisfying $\det D=\det \bar D$
can be always found, e.g.~the solution with $\det D=\det \bar D=1$
obtained by setting
$$
d_1:=\root n\of {\nu^-\over\hat \nu^-}\,,
\qquad
\bar d_1:=\root n\of {\hat \nu^+\over \nu^+}\,.
\eqno(4.13)$$
Thus we have a single orbit of $\Delta$ in this case.
If $n$ is even, then
$$
{\det D \over \det \bar D}= \left( {d_1\over \bar d_1}\right)^n
{\hat \nu^-\hat \nu^+\over \nu^- \nu_+}
\eqno(4.14)$$
can be set equal to $1$ by choosing $d_1$ and $\bar d_1$ if
and only if
$$
{\rm sign}\left(\nu^-\nu^+\right)={\rm sign}\left(\hat \nu^-
\hat \nu^+\right).
\eqno(4.15)$$
Hence in this case there are two orbits of $\Delta$ in the space
of pairs $(I_-,I_+)$ in correspondence with
${\rm sign}\left(\nu^-\nu^+\right)=\pm$. {\it Q.E.D.}
\medskip

For any integer $n>1$, let us introduce the matrices
$$
I_-^+:=\sum_{i=1}^{n-1} e_{i+1,i}
\quad\hbox{and}\quad
I_+^+:=\sum_{i=1}^{n-1} e_{i,i+1},
\quad
I_+^-:= \sum_{i=1}^{n-2} e_{i,i+1} - e_{n-1,n}.
\eqno(4.16)$$
It is clear from the proof of Proposition 1 that if $n$ is odd,
then any $(I_-, I_+)$  can be mapped  to $(I_-^+, I_+^+)$
by the action of the group $\Delta$.
If $n$ is even, then any $(I_-,I_+)$
can be mapped either  to $(I_-^+,I_+^+)$ or to $(I_-^+, I_+^-)$.
Thus,  for $n$ even, we may choose
$$
(\M^{\rm red}, \Omega^{\rm red}, \H^{\rm red})_\pm:=
(\M^{\rm red},\Omega^{\rm red}, \H^{\rm red})(I_-^+,I_+^\pm)
\eqno(4.17)$$
as representatives of the non-isomorphic reduced systems.
For  $n$ odd, we shall take the system
$(\M^{\rm red},\Omega^{\rm red}, \H^{\rm red})(I_-^+,I_+^+)$
as the representative.

Referring to the decomposition in (3.17), we can now describe
the set of Toda lattices contained in the reduced system as subsystems.

\medskip
\noindent
{\bf Proposition 2.} {\it
If $n$ is odd, then $(\M^{\rm red}, \Omega^{\rm red}, \H^{\rm red})$
contains all the possible Toda lattices, having any signs
for $\mu_i$, $i=1,\ldots, (n-1)$, in (4.1).
If $n$ is even, then the reduced system
$(\M^{\rm red}, \Omega^{\rm red}, \H^{\rm red})_\sigma$, $\sigma=\pm$,
contains
those Toda lattices for which the constants $\mu_i$
defining the Hamiltonian in (4.1) satisfy
$$
\sigma={\rm sign} \left( \prod_{i=1}^{n-1} \mu_i^{n-i}\right),
\eqno(4.18)$$
and each  of  these  Toda lattices  occur in two copies,
on the submanifolds $\M_{\pm m}^{\rm red}$
for $\pm m\in {\bf M}$ in (3.17).}

\medskip
\noindent
{\it Proof.}
According to the theorem of the preceding section,
the Bruhat decomposition of $SL(n,\Real )$ gives rise to the Toda
lattices
$
\left(\M^{\rm red}_{m},
\Omega^{\rm red}_{m}, \H^{\rm red}_{m}\right)
\subset (\M^{\rm red}, \Omega^{\rm red}, \H^{\rm red})$
having the Hamiltonian in (4.1) with $\mu_i=\nu_i^+ \nu_i^-  m_i m_{i+1}$
for any $m\in {\bf M}$ (cf.~(3.26)).
Using the above chosen representatives of the
non-isomorphic reduced systems, if $n$ is odd one finds that
$\mu_i=m_im_{i+1}$.
For arbitrarily given $\mu_i\in \{\pm 1\}$,
this has a unique solution for $m\in {\bf M}$,
which implies the claim.
On the other hand, when $n$ is even,
one has $\mu_i=m_i m_{i+1}$ for $\sigma = +$ and
$\mu_i=m_im_{i+1}$ $(i\neq n-1)$,
$\mu_{n-1}=-m_{n-1} m_n$ for $\sigma=-$.
The claim then follows by means
of a computation similar to that used in the proof of
Proposition 1.
{\it Q.E.D.}

\smallskip
\noindent
{\it  Remark 2.}
For  $n$ even, the systems
$(\M^{\rm red}, \Omega^{\rm red}, \H^{\rm red})_+$ and
$(\M^{\rm red}, \Omega^{\rm red}, \H^{\rm red})_-$ are not
isomorphic since only the former contains a Toda lattice
whose Hamiltonian is bounded from below.
Note also  that  the transformation given by
$q_i\mapsto -q_{n+1-i}$, $p_i\mapsto -p_{n+1-i}$ for any $i$
provides an  isomorphism between the Toda lattices
with  respective Hamiltonians
$H(q,p)={1\over 2}\tr(p^2)+\sum_{i=1}^{n-1} \mu_i e^{q_i-q_{i+1}}$
and
$H(q,p)={1\over 2}\tr(p^2) +\sum_{i=1}^{n-1} \mu_{n-i} e^{q_i-q_{i+1}}$.
(In the above we did not take this into account for counting the
Toda lattice content  of  $\M^{\rm red}$.)

\bigskip
\medskip
\centerline{\bf 5.~An involutive symmetry of the reduced system}
\medskip

Below we exhibit an involutive symmetry of the reduced
system, which is induced by a corresponding symmetry of the
original system on $\M=T^* SL(n,\Real)$.
In the final analysis,
this symmetry is due
to the reflection symmetry of
the Dynkin diagram of the Lie algebra $sl(n,\Real)$, and it
may be thought of as a global version of the symmetry
mentioned at the end of Section 4.
It will be used to simplify some arguments later in the paper.

Denote by $X^\tau$ the transpose of any $n\times n$ matrix $X$
 with respect to the {\it anti-diagonal},
$$
(X^\tau)_{i,j}:=X_{n+1-j, n+1-i}\,,
\qquad
\forall\, 1\leq i,j\leq n.
\eqno(5.1{\rm a})$$
This operation has similar properties as the usual transpose since
$$
X^\tau=\eta X^T\eta^{-1}\quad\hbox{with}\quad \eta_{i,j}:=\delta_{i,n+1-i}.
\eqno(5.1{\rm b})$$
Then define the transformation $\varphi:\M\rightarrow \M$  by
$$
\varphi: (g,J)\mapsto ( (g^{-1})^\tau, -J^\tau ),
\qquad
\forall\, (g,J)\in \M.
\eqno(5.2)$$
Clearly, $\varphi$  is an {\it involutive  symmetry} of the Hamiltonian
system $(\M,\Omega,\H)$.
On the constrained manifold defined in (3.11a), it acts according  to
$$
\varphi: \M^{\rm c}(I_-,I_+)\rightarrow
\M^{\rm c}(-I_-^\tau, -I_+^\tau).
\eqno(5.3)$$
Since
$$
n_\pm^\tau\in N_\pm\quad\hbox{for}\quad  n_\pm\in N_\pm\,,
\eqno(5.4)$$
$\varphi$ in (5.3) maps gauge orbits to gauge orbits.
Hence it naturally induces  a mapping
$$
\hat \varphi : \M^{\rm red}(I_-,I_+)\rightarrow
\M^{\rm red}(-I_-^\tau, -I_+^\tau),
\eqno(5.5)$$
which is an {\it isomorphism} between the corresponding reduced
systems.
On the other hand, a straightforward calculation shows  that
the pairs $(I_-, I_+)$ and $(-I_-^\tau, -I_+^\tau)$ can
be transformed into each other by the action of the group
$\Delta$, i.e., there exist matrices $(D,\bar D)$ of the form in (4.2)
satisfying
$$
I_-=D (-I_-^\tau) D^{-1},
\qquad
I_+=\bar D (-I_+^\tau) \bar D^{-1}.
\eqno(5.6)$$
The  corresponding symmetry $\phi_{(D,\bar D)}$ given in (4.3)
induces the isomorphism
$$
\hat \phi_{(D,\bar D)}: \M^{\rm red}(-I_-^\tau,-I_+^\tau)\rightarrow
\M^{\rm red}(I_-, I_+).
\eqno(5.7)$$
Composing the above two mappings, we  have the involutive
symmetry
$$
\psi_{(D,\bar D)}:=\phi_{(D,\bar D)}\circ \varphi :\M\rightarrow \M
\eqno(5.8{\rm a})$$
given explicitly as
$$
\psi_{(D,\bar D)}:
(g,J)\mapsto ( D(g^{-1})^\tau \bar D^{-1}, -D J^\tau D^{-1}),
\qquad
\forall\, (g,J)\in \M.
\eqno(5.8{\rm b})$$
The involution property,
$\psi_{(D,\bar D)}^2={\rm id}$, is easily verified and
$\psi_{(D,\bar D)}$ maps the constrained manifold
$\M^{\rm c}(I_-,I_+)$ to itself.
The construction implies the following proposition.

\medskip
\noindent
{\bf Proposition 3.}
{\it
Fix a  pair $(I_-,I_+)$ according to (2.2).
Then $\psi_{(D,\bar D)}$ given in (5.8) induces the mapping
$$
\hat \psi_{(D,\bar D)}=\hat \phi_{(D,\bar D)}\circ \hat \varphi:
\M^{\rm red}(I_-,I_+)\rightarrow \M^{\rm red}(I_-,I_+),
\eqno(5.9)$$
which is an involutive symmetry of the reduced system
$(\M^{\rm red}, \Omega^{\rm red}, \H^{\rm red})(I_-,I_+)$.
}
\medskip

It is worth noting that  $\hat \psi_{(D,\bar D)}$ permutes
the Toda lattices associated with the Bruhat decomposition,
$$
\hat \psi_{(D,\bar D)}: \M^{\rm red}_m\rightarrow \M^{\rm red}_{m'}
\quad\hbox{with}\quad
m'_{i}={\rm sign}(d_i)\, m_{n+1-i}\, {\rm sign}(\bar d_i),
\eqno(5.10)$$
where $d_i:=D_{i,i}$ are the entries of the diagonal matrix  $D$,
and similarly for $m$, $m'$ and $\bar D$.
The mapping in (5.10)  can be recognized to be just the isomorphism between
the respective Toda lattices  remarked in  Section 4.

We now  describe the symmetry transformation $\psi_{(D,\bar D)}$ explicitly.
We consider the case for which $I_-=I_-^+$, $I_+=I_+^+$ in (4.16),
since this will be used later.
Defining the $n\times n$ diagonal matrix $D_+$ by
$$
(D_+)_{i,j}:=(-1)^i\delta_{i,j}, \qquad \forall\, 1\leq i,j\leq n,
\eqno(5.11)$$
we may take
$$
(D,\bar D)=(D_+, D_+)
\eqno(5.12)$$
as the solution of (5.6).
Up to the trivial redefinition $(D,\bar D)\mapsto (-D, -\bar D)$,
which does not change $\psi_{(D,\bar D)}$, this is the unique solution
if $n$ is odd.
If $n$ is even, then
$$
(D,\bar D)=(D_+,-D_+)
\eqno(5.13)$$
is another solution.
In this case, the  composed transformation
 $\chi:=\psi_{(D_+,D_+)}\circ \psi_{(D_+, -D_+)}$
simply operates as
$$
\chi: (g,J)\mapsto (-g,J),
\qquad
\forall (g,J)\in \M.
\eqno(5.14)$$
In particular, the symmetry
$\hat \chi: \M^{\rm red}\rightarrow \M^{\rm red}$
induced by $\chi$ is responsible for the
occurrence  of the Toda
lattices in the reduced system  in two copies (cf.~Proposition 2).
This symmetry is available only for even $n$,
because $\det(-g)=(-1)^n$.

Let us now  consider the set of functions $Q_i$
on $\M=G\times \G$ given by the principal minors of the matrix
$g\in G=SL(n,\Real)$,
$$
Q_i(g,J):=\bar Q_i(g),
\qquad
\bar Q_i(g):=\det\pmatrix{g_{i,i}&\cdots &g_{i,n}\cr
                          \vdots&{}&\vdots\cr
                           g_{n,i}&\cdots &g_{n,n}\cr}
\quad\hbox{for}\quad i=2,\ldots,n.
\eqno(5.15)$$
These functions are {\it invariant} under our gauge group $N_+\times N_-$,
$$
\bar Q_i(n_+ g n_-^{-1})=\bar Q_i(g),
\qquad
\forall\, n_\pm\in N_\pm,\quad  i=2,\ldots, n.
\eqno(5.16)$$
On each connected component $G_m$ of the big cell in (3.14),
the signs of the functions $\bar Q_i$
are  fixed according to
$$
{\rm sign}\left( \bar Q_i\right)\vert_{G_m}=
{\rm sign}\left( \prod_{k=i}^n m_k\right),
\eqno(5.17)$$
and these functions  are in one-to-one correspondence with
the components of $q$ appearing in
the decomposition $g=n_+ me^q n_-$ for $g\in G_m$.
The importance of the $Q_i$ is that, unlike the components
of $q$, they give rise  to globally defined smooth
functions\footnote{${}^{3}$}{\ninerm
Intuitively, the functions on $\M^{\rm red}$
corresponding to the $Q_i$ can be thought of  as global ``position''
variables. But one must be careful since it appears (see Sections 6 and 7)
 that in general $\M^{\rm red}$ is {\it not} a cotangent bundle of some
configuration space.}
on $\M^{\rm red}$.

For later reference, we  present the behaviour
of the $Q_i$ with respect to the symmetry transformation
$\psi_{(D_+, D_+)}$ given above.

\medskip
\noindent
{\bf Proposition 4.} {\it
 Using the above notations,
we have $Q_i \circ \psi_{(D_+,D_+)}=Q_{n+2-i}$ for any
$i=2,\ldots, n$, that is,
$$
\bar Q_i\left(D_+ (g^{-1})^\tau D_+^{-1}\right)
=\bar Q_{n+2-i}\left(g\right),
\quad
\forall g\in SL(n,\Real).
\eqno(5.18)$$}
\noindent
{\it Proof.}
The equality  $\left(D_+ g D_+^{-1}\right)_{i,j}=(-1)^{i+j} g_{i,j}$
implies that
$\bar Q_i\left( D_+ g D_+^{-1}\right) =Q_i(g)$ for any $i=2,\ldots, n$.
Hence it is enough to show that
$$
\bar Q_i\left((g^{-1})^\tau \right)=\bar Q_{n+2-i}(g),
\qquad
\forall g\in SL(n,\Real).
\eqno(5.19)$$
 For  an element $g=n_+ me^q n_-$  of the big cell
of $SL(n,\Real)$, because  of (5.16), (5.19) is equivalent to
$$
\bar Q_i\left(e^{-{q^\tau} }m^\tau\right)=\bar Q_{n+2-i}\left(me^q\right),
\eqno(5.20)$$
which is readily verified.
This completes the proof, since the  big cell is
a dense submanifold of  $SL(n,\Real)$.
{\it Q.E.D.}
\medskip

To summarize, at this stage we have a clear understanding of
the Toda lattice content of the non-equivalent reduced systems
and their residual discrete symmetries.
Next we take some initial steps towards
 investigating the global structure
of these systems.

\bigskip
\medskip
\centerline{\bf 6.~A  hypersurface model of
$\M^{\rm red}$ from global gauge fixing}
\medskip

In section 3 we used a
{\it locally} --- over the big cell of $SL(n,\Real )$ ---
valid gauge fixing to identify the reduced system
$(\M^{\rm red},\Omega^{\rm red}, \H^{\rm red})$ obtained from
$T^* SL(n,\Real )$ as one containing $2^{n-1}$ Toda lattices
glued together along lower dimensional submanifolds.
To furnish  a tool for studying the
topology of $\M^{\rm red}=\M^{\rm c}/N$,
 we here describe a {\it global}
 cross section of the gauge orbits in $\M^{\rm c}$.
The global gauge fixing permits us to find a model
of the manifold $\M^{\rm red}$ in the form of a hypersurface in
${\Real}^{2n-1}$.

The constraints on $J$ given in (3.11) are  well studied
in the context of generalized KdV equations
and $\cal W$-algebras.
Drinfeld and Sokolov has shown in \q{\DS} how to define  a global 
gauge fixing
for the gauge transformations generated by these constraints.
The corresponding gauges  are called ``DS gauges'', e.g., in \q{\FORTW}.
To obtain a global gauge fixing for the constraints in (3.11),
we simply have to restrict both $J$ and $\tilde J$ to a DS gauge.
This has to be formulated in terms of the variables
$(g,J)$ since our phase space is not $\G\times \G$ but
$\M=T^*G\simeq G\times \G$, with $G=SL(n,\Real )$ and $\G=sl(n,\Real )$.
The so obtained global cross section of the $N$-orbits in $\M^{\rm c}$ 
will be henceforth called a ``double DS gauge'' (see also \q{\TF}).

A  double DS gauge can be described  as follows.
Let $V\subset \G_{\geq 0}$ and $\tilde V\subset \G_{\leq 0}$,
where we use the principal
grading
of $\G$ as in (2.1),
be graded linear subspaces\footnote{${}^{4}$}{\ninerm
That is, $V$ and $\tilde V$
are invariant subspaces of
${\rm ad\,} I_0$ for $2I_0={\rm diag}\left(n-1, n-3, \ldots, 1-n\right)$.}
appearing in a
linear direct sum decomposition
$$
\G_{\geq 0}=[I_-, \G_{>0}]+V,
\qquad
\G_{\leq 0}=[I_+, \G_{<0}] + \tilde V.
\eqno(6.1)$$
Then define $\M^{\rm DS}\subset \M^{\rm c}$ as
$$
\M^{\rm DS}=\left\{\, (g,J)\in \M^{\rm c}\,\vert\, J\in (I_- + V),
\quad (g^{-1} Jg)\in (I_+ +\tilde V)\,\right\}.
\eqno(6.2)$$
The point is that this is a global cross section of the gauge orbits for
any choice of $V$, $\tilde V$, and hence
we may identify $\M^{\rm red}=\M^{\rm c}/N= \M^{\rm DS}$.
Using (2.2), a convenient choice of $V$, $\tilde V$ is furnished by
$$
V:={\rm span}\left\{\, e_{1,i}\ \,  \hbox{for}\ \, i=2,\ldots, n\,\right\},
\qquad
\tilde V:=
{\rm span}\left\{\,e_{i,1}\ \, \hbox{for}\ \, i=2,\ldots,n\,\right\}.
\eqno(6.3)$$
Corresponding to this choice, $(g,J)\in \M^{\rm DS}$
is restricted to satisfy
$$
J=I_- + \sum_{i=2}^n u_i e_{1,i},
\qquad
g^{-1} J g=I_+ + \sum_{i=2}^n v_ie_{i,1},
\quad\hbox{with some}\quad u_i, v_i\in \Real.
\eqno(6.4)$$
Notice that the parameters $v_i$ can be determined in
terms of the $u_i$ from the relations
$$
\tr\, J^k =\tr\left(g^{-1} J g\right)^k,
\qquad
\forall\, k=2,\ldots,n.
\eqno(6.5)$$
Our problem is to solve these relations and to find a proper
parametrization of $(g,J)\in \M^{\rm DS}$.

The solution of (6.5) depends on the choice of $(I_-,I_+)$.
Choosing  the representatives in (4.16),
we first treat the system
$$
(\M^{\rm red}, \Omega^{\rm red},\H^{\rm red})(I_-^+, I_+^+).
\eqno(6.6)$$
{}From (6.4), (6.5) and  $(I_+^+)^T=I_-^+$, we have
$$
\tr J^k= \tr (I_-^++\sum_{i=2}^n u_i e_{1,i})^k
=\tr (g^{-1} Jg)^k
=\tr ( (g^{-1} Jg)^T )^k=
\tr (I_-^++\sum_{i=2}^n v_i e_{1,i})^k,
\eqno(6.7)$$
which implies
$$
v_i=u_i,\qquad \forall\, i=2,\ldots, n.
\eqno(6.8)$$
Then the second relation in (6.4) becomes,
$$
( I_-^+ + \sum_{i=2}^n u_i e_{1,i}) g=g(I_+^+ + \sum_{i=2}^n u_i e_{i,1}).
\eqno(6.9)$$
It follows from (6.9)  that
the matrix $g$ has the same value along each
anti-diagonal line,
$$
g_{i,j} = g_{k,l} \qquad \hbox{if} \quad i + j = k + l.
\eqno(6.10)
$$
It also follows that
$$
g_{1,j-1} = \sum_{i = 2}^n u_i g_{i,j}\,, \qquad \forall\,  j 
= 2, \ldots, n,
\eqno(6.11)$$
which shows that
all the ``lower'' entries
$g_{1,j-1}$ in the first row for $1 \leq j-1 \leq n-1$
can be expressed in terms of the
``higher'' ones together with the $u_i$ for $2 \leq i \leq n$.
Thus, if we set
$$
u_{i+j} = g_{i,j}  \qquad \hbox{for} \quad
  n+1 \leq i + j \leq 2n,
\eqno(6.12)$$
then
in the double DS gauge
the matrix $g$ can be written as
$$
g = \pmatrix{
\,g_{1,1}(u)\hfill & g_{1,2}(u)\hfill & \ldots
                  & g_{1,n-1}(u)\hfill & u_{n+1}\hfill \cr
\,g_{1,2}(u)\hfill & g_{1,3}(u)\hfill & \ldots
                  & u_{n+1}\hfill      & u_{n+2}\hfill \cr
\,\, \vdots\hfill &
     \, \vdots\hfill &&
        \, \vdots\hfill &
           \, \vdots\hfill &\cr
\,g_{1,n-1}(u)\hfill & u_{n+1}\hfill & \ldots
                    & u_{2n-2}\hfill     & u_{2n-1}\hfill \cr
\,u_{n+1}\hfill      & u_{n+2}\hfill & \ldots
                    & u_{2n-1}\hfill     & u_{2n}\hfill   \cr
}.
\eqno(6.13)$$
In this way  the set of variables $(u_2, u_3,\ldots, u_{2n})$
provides a parametrization of $(g,J)\in \M^{\rm DS}$.
These $(2n-1)$ parameters are subject to the relation
$\det g(u)=1$ in accordance with the fact that
the dimension of $\M^{\rm c}/N$ is $2(n-1)$.

\medskip
\noindent
{\it Remark 3.} For any choice of $V$, $\tilde V$ in (6.1),
the double DS gauge $\M^{\rm DS}\subset \M$ in (6.2) is invariant
with respect to the  original dynamics on $\M$ since
$\M^{\rm DS}$ is defined by constraining $J$ and $\tilde J$,
which are constants of motion.
Using the model $\M^{\rm red}=\M^{\rm DS}$,
the trajectory $(g(t), J(t))\in \M^{\rm DS}$
of the reduced Hamiltonian system
associated with the initial value $(g,J)\in \M^{\rm DS}$ at $t=0$
is given by
$$
J(t)=J, \qquad g(t)=e^{tJ} g.
\eqno(6.14)$$
For our choice of $V$, $\tilde V$ in (6.3),
the evolution
equation, ${d J(t)\over dt}=0$, ${d g(t)\over dt}=J(t)g(t)$, of the reduced
system in (6.6) can be re-casted  as
$$
{d u_i(t)\over dt}=0, \quad\forall\, i=2,\ldots,n,
\qquad
u_k(t) ={d^{2n-k} u_{2n}(t)\over dt^{2n-k}}\,,
\quad\forall\, k=n+1,\ldots, 2n-1,
\eqno(6.15{\rm a})$$
and
$$
{d^n u_{2n}(t)\over dt^n}-\sum_{i=2}^n 
u_i {d^{n-i} u_{2n}(t)\over dt^{n-i}}=0.
\eqno(6.15{\rm b})$$
Solving  this linear differential equation with  {\it constant}
coefficients  for a given  initial value $(u_2,\ldots, u_{2n})$,
subject to $\det g(u)=1$ with $g(u)$ in (6.13),
is equivalent  to computing  the last row of the matrix
$e^{tJ}$ in (6.14).

\medskip

We wish to present a reinterpretation of  the above
identification $\M^{\rm red}=\M^{\rm DS}$.
For  this let us now think of $u:=(u_2, u_3,\ldots, u_{2n})$ as the general
element of the manifold $\Real^{2n-1}$ and define
the polynomial $P$ on  $\Real^{2n-1}$ by
$$
P: u\mapsto \det g(u),
\eqno(6.16)$$
where $g(u)$ is determined by (6.13) using  (6.10), (6.11).
We  then define
$$
{\cal S}_+(n):=\{\, u\in \Real^{2n-1}\,\vert\, P(u)=1\,\}.
\eqno(6.17)$$
Computations made for small $n$ (see the next section)
support the following {\bf conjecture}:
$$
dP(u)\vert_{P(u)=1}\neq 0.
\eqno(6.18)$$
If (6.18) is valid,  then ${\cal S}_+(n)\subset \Real^{2n-1}$
is a {\it regular} submanifold.
On the other hand, we can consider the mapping
$$
(u_2,u_3, \ldots, u_{2n}):  \M^{\rm DS} \rightarrow \Real^{2n-1}
\eqno(6.19)$$
engendered  by the above parametrization of $(g,J)\in \M^{\rm DS}$.
This mapping is smooth by construction.
In fact, it follows from DS gauge fixing
that $u_2,u_3, \ldots, u_{2n}$ are given by polynomials
when regarded as gauge invariant functions on $\M^{\rm c}$.
It is also a tautological statement that
the mapping in (6.19) is  a one-to-one mapping from
$\M^{\rm DS}$ to its image given by ${\cal S}_+(n)\subset \Real^{2n-1}$.
If ${\cal S}_+(n)\subset \Real^{2n-1}$ is a regular submanifold,
then we can conclude from
a well-known theorem in differential geometry that
the mapping in (6.19) gives rise to a {\it diffeomorphism}
$$
(u_2, u_3,\ldots, u_{2n}): \M^{\rm DS} \rightarrow {\cal S}_+(n).
\eqno(6.20)$$
This is a non-trivial statement since
the manifold structure of $\M^{\rm DS}=\M^{\rm c}/N$ is determined
by the reduction  and that of ${\cal S}_+(n)$ is determined
by its embedding into $\Real^{2n-1}$.

In conclusion, modulo the conjecture given by  (6.18),
we have shown that {\it the manifold $\M^{\rm red}(I_-^+,I_+^+)$
is diffeomorphic to the regular hypersurface
${\cal S}_+(n)\subset \Real^{2n-1}$ for any $n>1$}.

\medskip

We can analogously treat the system
$(\M^{\rm red}, \Omega^{\rm red},\H_{\rm red})_-$ defined for any
even $n$ by taking  $(I_-, I_+)=(I_-^+, I_+^-)$ in (4.17).
Making use of the relation
$$
I_+^-=D_0 I_+^+ D_0^{-1}
\quad\hbox{with}\quad
D_0={\rm diag}(1,\ldots,1,-1),
\eqno(6.21)$$
instead of (6.8) we  find
$$v_i=u_i
\quad\hbox{for}\quad  i=2,\ldots, (n-1)
\quad\hbox{and}\quad  v_n=-u_n.
\eqno(6.22)$$
Denoting the general element of $\M^{\rm DS}$
now as $(\bar g,J)\in \M^{\rm DS}$, it turns out  that
$$
\bar g =g D_0,
\eqno(6.23)$$
where the matrix $g=g(u)$ is still given by the same
equations (6.10) through (6.13).
The only difference is that in this case, due to (6.23),
the restriction on the parameters $u=(u_2,u_3,\ldots, u_{2n})$  reads
$$
P(u)=\det g(u)=-1.
\eqno(6.24)$$
Similarly to (6.18), we have the {\bf conjecture}:
$$
dP(u)\vert_{P(u)=-1}\neq 0.
\eqno(6.25)$$
If this holds, then  $\M^{\rm DS}$ is diffeomorphic
to the regular hypersurface
$$
{\cal S}_-(n):=\{\, u\in \Real^{2n-1}\,\vert\, P(u)=-1\,\}.
\eqno(6.26)$$
The regularity conjecture of equations  (6.18), (6.25) will be
seen to hold for the examples of  the next section.

\bigskip
\medskip
\centerline{\bf 7.~ Examples}
\medskip

Below we develop the simplest examples of the construction
of  the preceding sections by studying the reduced systems
$(\M^{\rm red}, \Omega^{\rm red}, \H^{\rm red})$
resulting from $T^* SL(n,{\Real})$ for $n=2,3$.
The conjectures in (6.18), (6.25) will be verified for  these  examples,
which will illustrate
the regularization of the singular Toda lattices.
We shall also show that the Toda lattice having the Hamiltonian
$H(q,p)={1\over 2}\tr ( p^2) + \sum_i e^{q_i-q_{i+1}}$,
for which regularization is not necessary since the Hamiltonian
vector field is complete,
is realized as a {\it connected  component} of the reduced phase space.
We  conjecture that this  Toda
lattice with good sign --- which is contained in
$(\M^{\rm red}, \Omega^{\rm red}, \H^{\rm red})(I_-^+,I_+^+)$
according to Proposition 2 ---  is always topologically
disconnected from the rest of the reduced phase space.
It would be interesting to prove (or disprove) this conjecture for
general $n$, but with our ``direct inspection''
method this is  difficult as can be seen from Appendix A,
where the conjecture is  verified for $n=4$.

\medskip
\medskip
\noindent{\bf 7.1.~The case of $SL(2,{\Real})$}
\medskip

We wish to  treat both  systems
$$
(\M^{\rm red}, \Omega^{\rm red}, \H^{\rm red})_\sigma
\quad\hbox{for}\quad \sigma=\pm\,
\eqno(7.1)$$
defined in (4.17).
The matrices  $J(u)$ of (6.4) and  $g(u)$ of (6.12) can be written
$$
J(u)=\pmatrix{0&u_2\cr 1&0\cr},
\qquad
g(u)=\pmatrix{u_2u_4&u_3\cr u_3&u_4\cr}.
\eqno(7.2)$$
Hence the equation of the hypersurface
${\cal S}_\sigma(2)\subset {\Real}^3$ is given by
$$
P(u)=\det g(u)=u_2 u_4^2 -u_3^2 =\sigma 1.
\eqno(7.3)$$
It is straightforward to check that
$$
dP(u)\vert_{P(u)=\sigma 1}\neq 0,
\eqno(7.4)$$
which implies that ${\cal S}_\sigma(2)\subset \Real^3$ is
a regular hypersurface diffeomorphic to $\M^{\rm red}_\sigma$.
Regarding the $u_i$ as gauge invariant functions on the
constrained manifold $\M^{\rm c}_\sigma$, we find the Poisson brackets
$$
\{ u_4, u_3\}={u_4^2\over 2},
\qquad
\{ u_4,u_2\}=u_3,
\qquad
\{ u_3, u_2\}=u_4 u_2.
\eqno(7.5)$$
This fixes the reduced  dynamics since the Hamiltonian is simply
$$
\H^{\rm red}_\sigma(u)={1\over 2}\tr J^2(u)=u_2.
\eqno(7.6)$$

\noindent
{\it Remark 4.}
One may  use (7.5) to define
a Poisson bracket on ${\Real}^3$ with coordinates $(u_2,u_3,u_4)$.
In this case  $P(u)$ in (7.3) acquires the interpretation of  a
Casimir function on the Poisson manifold $({\Real}^3, \{\ ,\ \})$.
Then ${\cal S}_\pm(2)\subset \Real^3$ are  symplectic leaves.
\medskip

In the  $\sigma=+$ case equation (7.3) excludes the value $u_4=0$
and the surface ${\cal S}_+(2)$ is easily seen to be the union
of two disconnected pieces, which are diffeomorphic
to the plane as  they can be parametrized respectively by
$(u_3, u_4)\in \Real\times \Real^+$ and by 
$(u_3,u_4)\in \Real\times \Real^-$.
On these two coordinate patches we can introduce new coordinates
$(x_\pm, \pi_\pm)\in {\Real}^2$ by
$$
u_4=\pm \exp(-x_\pm ),
\quad
u_3=-\pi_\pm u_4 ,
\eqno(7.7)$$
in terms of which the reduced Poisson brackets is given by
$\{ x_\pm, \pi_\pm \}={1/2}$
and the reduced Hamiltonian, $\H^{\rm red}_+$, is written as
$$
\H^{\rm red}_{+}(x_\pm, \pi_\pm) = \pi_\pm^2 + \exp(2x_\pm).
\eqno(7.8)$$
These relations take the form of equations (2.4), (2.5)  in terms of
the respective variables $q_\pm={\rm diag}( x_+, -x_+)$ and
$p_\pm ={ \rm diag}(\pi_+, -\pi_+)$.

In the above  we have decomposed  the reduced system into
the disconnected union of two Toda lattices with good sign.
Of course these subsystems arise from the components of the big cell of
$SL(2,\Real )$ containing $\pm e$.
Since $u_4$ is nothing but the gauge invariant
function of $(g,J)\in \M^{\rm c}$ given by the matrix element $g_{22}$,
the inequality $u_4=g_{22}\neq 0$  excludes
the part of $T^* SL(2,{\Real} )$ that does not lie over
the big cell.
This is the reason for the disconnectedness of
$\M^{\rm red}_{\sigma=+}={\cal S}_+(2)$.

Let us now discuss  the case of $\sigma =-$.
This is drastically different from the previous case, because
equation (7.3) no longer  excludes  the value $u_4=0$
corresponding to the complement of the big cell of $SL(2,\Real)$.
We still have  two Toda lattices as subsystems  for
$u_4>0$ and for $u_4<0$, but now the reduced Hamiltonian is given
for these subsystems by
$$
\H^{\rm red}_{-}(x_\pm, \pi_\pm) = \pi_\pm^2 - \exp(2x_\pm).
\eqno(7.9)$$
All trajectories of these subsystems  are
singular, since in fact they reach infinity at finite time.
But there is no singularity at all in the full reduced system
$(\M^{\rm red}, \Omega^{\rm red}, \H^{\rm red})_-$, whose trajectories
are simply the connected components of the curves
obtained by intersecting ${\cal S}_-(2)\subset \Real^3$ with the planes of
fixed
energy $u_2=E$ for any constant $E$.
These curves are hyperbolae or ellipses depending on the sign of $E$,
$$
u_3^2 - E u_4^2 =1.
\eqno(7.10)$$
At negative energy $E<0$ the trajectory is periodic.
The  corresponding
closed geodesic on  $SL(2,\Real)$ connects the two components
of the big cell, ${\rm sign}(u_4)=\pm$, through the lower
dimensional submanifold, $u_4=0$.
In effect, the two  singular Toda
lattices  are  regularized by their
trajectories being glued  together ``at infinity''.
This  illustrates the regularization obtained for
general $n$  as explained at the end of Section 3.
It is worth noting that the surface ${\cal S}_-(2)$ is
{\it not simply connected}. In fact, the loop given by,
say, the ellipse in
(7.10) for some $E<0$ cannot be contracted to a point on
the  surface.

\medskip
\medskip
\noindent{\bf 7.2.~The case of $SL(3,{\Real})$}
\medskip

We next  analyze  the structure of the reduced phase space
then discuss the integration of the equation of motion.

\noindent
$\underline{\hbox{\it i) The reduced phase space
for $SL(3,{\Real} )$.}}$
We can choose $I_-=I_-^+$, $I_+=I_+^+$ without loss of generality
since all choices of $(I_-, I_+)$ are equivalent for $SL(n,\Real)$
with odd $n$.
Then the general element $(g,J)\in \M^{\rm DS}$ can be parametrized as
$$
J=\pmatrix{0&u_2&u_3\cr1&0&0\cr0&1&0\cr}
\qquad\hbox{and}\qquad
g=\pmatrix{(u_2 u_4 +u_3u_5)&(u_2 u_5 +u_3 u_6)&u_4\cr
(u_2u_5+u_3 u_6)&u_4&u_5\cr
u_4&u_5&u_6\cr}.
\eqno(7.11)$$
In order to prove  the regularity of the hypersurface
${\cal S}_+(3)\subset \Real^5$ defined by the equation
$$
P(u):=\det g(u)=1,
\eqno(7.12)$$ we now verify that $dP$ does not vanish
at any point of ${\cal S}_+(3)$.
For this it will be enough to inspect  the vanishing
of the first three components of $dP$,
$$\eqalign{
{\pa P\over \pa u_2}&=(u_4^2u_6+u_4u_5^2)-2u_5u_6(u_2u_5+u_3u_6)=0,\cr
{\pa P\over \pa u_3}&=(3u_4u_5u_6-u_5^3)-2u_6^2(u_2u_5+u_3u_6)=0,\cr
{\pa P\over \pa u_4}&=u_2(2u_4u_6+u_5^2)+u_3(3u_5u_6)-3u_4^2=0.\cr}
\eqno(7.13)$$
To show that $dP=0$ and $P=1$ are not compatible,
we proceed by distinguishing various cases.
First we look at the case where we assume that
$$
u_5u_6\neq 0.
\eqno(7.14)$$
Then we can combine the first two equations in (7.13) to yield
$$
u_4={u_5^2\over u_6}.
\eqno(7.15)$$
Substituting this back into the first equation in (7.13) gives
$$
u_2={u_5^2\over u_6^2} -u_3{u_6\over u_5}.
\eqno(7.16)$$
Actually, the third equation in (7.13) is then also satisfied.
However, if we now plug back (7.15), (7.16) into
(7.12) we find that the determinant vanishes since
its rows are all proportional with each other.
Therefore we have proved that (7.14) is not compatible
with the requirement that  $P=1$ and $dP=0$.
So a ``singularity'' can only occur at a point where (7.14)
does not hold.
There are then 3 further sub-cases, such as
$$
u_5=u_6=0,
\quad\hbox{or}\quad
u_6= 0,
\quad\hbox{or}\quad
u_5=0.
\eqno(7.17)$$
The reader can easily check that
these are all excluded  by     (7.12) and (7.13).
In conclusion, we have proved that ${\cal S}_+(3)\subset \Real^3$ is a
{\it regular hypersurface}.
This implies that ${\cal S}_+(3)$ is  {\it diffeomorphic}
to $\M^{\rm red}$ as was explained in Section 6.

\smallskip

We know from Proposition 2 that the reduced system
$(\M^{\rm red}, \Omega^{\rm red}, \H^{\rm red})$
contains the  Toda lattice with good sign.
For   $(I_-,I_+)=(I_-^+, I_+^+)$,  using
the  Bruhat decomposition in (3.17),
this Toda lattice lives on the open submanifold
$\M^{\rm red}_e\subset \M^{\rm red}$.
To identify $\M^{\rm red}_e$  in terms of  the double DS gauge
in (7.11), we consider the mapping
$$
Q:=(Q_2,Q_3): \M^{\rm red}\rightarrow \Real\times \Real,
\qquad
Q_3(g,J)= u_6, \quad
Q_2(g,J)=\det\pmatrix{u_4&u_5\cr u_5&u_6\cr}.
\eqno(7.18)$$
Since $g$ is Gauss decomposable  if an only if
its  principal minors are positive, we have
$$
\M^{\rm red}_e=Q^{-1}\left(\Real^+\times \Real^+\right).
\eqno(7.19)$$

\smallskip
\noindent
{\bf Proposition 5.}
{\it The phase space $\M^{\rm red}_e$
of the $SL(3,\Real )$ Toda lattice with good sign
is a connected  component of $\M^{\rm red}$, i.e.,
it is disconnected from its complement.}
\smallskip

\noindent
{\it Proof.}
The claim  that the open submanifold
$\M^{\rm red}_e\subset \M^{\rm red}$ is disconnected from
its complement follows  if for some arbitrary $t_0\in \Real$ and
$\epsilon >0$
we demonstrate the {\it non-existence} of a
continuous path,  $\gamma(t)=(g(t), J(t))\in \M^{\rm red}$
for $t\in  [t_0, t_0+\epsilon ]$,   such that
$$
Q_2(\gamma(t))>0,\quad Q_3(\gamma(t))>0  \qquad \hbox{for}\quad t>t_0
\eqno(7.20)$$
and
$$
Q_2(\gamma(t_0))Q_3(\gamma(t_0))=0.
\eqno(7.21)$$
To give an indirect proof, suppose that
the statement is not true.
Then there exists
a  continuous  path  $\gamma(t)$
satisfying (7.20) and (7.21).
Let us assume for the moment
that (7.20) and (7.21) imply
$$
Q_2(\gamma(t_0))=Q_3(\gamma(t_0))=0.
\eqno(7.22)$$
Then  we see from (7.18)  that
$$
u_5(t_0)=0.
\eqno(7.23)$$
Looking at $g(u)$ in (7.11) at $t=t_0$,
we also  see that the condition $\det g(u) =1$ requires
$$
u_4(t_0)=-1.
\eqno(7.24)$$
However, (7.20) and the formula of $Q_2$ imply that
$$
u_4 (t)>0
\quad\hbox{for}\quad t>t_0.
\eqno(7.25)$$
If the curve $\gamma(t)$ is continuous
then $u_4 (t)$ is a continuous function at $t=t_0$,
which is impossible by the above two equations.

To complete the proof, it remains to show (7.22).
Obviously, the other possibilities allowed by (7.21)
are
$$
Q_3(\gamma(t_0))=0
\quad\hbox{and}\quad
Q_2(\gamma(t_0))>0
\eqno(7.26)
$$
or
$$
Q_3(\gamma(t_0))>0
\quad\hbox{and}\quad
Q_2(\gamma(t_0))=0.
\eqno(7.27)$$
The first possibility (7.26) can be denied at once since
it is inconsistent with the definitions for $Q_2$ and
$Q_3$ (cf.(7.18)).
To deny (7.27), we notice  that,
according to Proposition 4,  the involutive symmetry
$\psi_{(D_+,D_+)}$ would send a curve satisfying
(7.20) and (7.27) into one satisfying (7.20) and (7.26),
because in the $SL(3,\Real)$  case $\psi_{(D_+,D_+)}$ interchanges
the minors $Q_2$ and $Q_3$.
Alternatively, we may use
$0 = Q_2(\gamma(t_0)) = u_4(t_0) u_6(t_0) - u_5^2(t_0)$
to obtain
$$
\eqalignno{
0 < Q_3^3(\gamma(t_0)) \det g(t_0) &=
    u_6^3(t_0) \det
   \pmatrix{ g_{11}  &   g_{12} &  u_4 \cr
             g_{12}  &      u_4 &  u_5 \cr
              u_4    &      u_5 &  u_6 \cr} (t_0)  \cr
    &= - \bigl\{ u_5^3(t_0) - g_{12}(t_0)
                    u_6^2(t_0)\bigr\}^2 \ ,
&(7.28)
}
$$
which is, again,
a contradiction. Thus (7.22) holds and the proof is complete.
{\it Q.E.D.}

\smallskip
\noindent
{\it Remark 5.}
Notice from the above proof that  the boundary of the positive
quadrant of $\Real \times \Real$  does not belong to the image
of the map  $(Q_2,Q_3):\M^{\rm red}\rightarrow \Real\times \Real$
induced  by the principal minors of $g\in SL(3,\Real)$.

\medskip
\noindent
$\underline{\hbox{\it ii) Trajectories of the reduced system
for $SL(3,{\Real} )$.}}$
In addition to the connected  component
$\M^{\rm red}_e$,  the system 
$(\M^{\rm red},\Omega^{\rm red}, \H^{\rm red})$
contains three Toda lattices  for which the  Hamiltonian is
not bounded from below.
We below exhibit   trajectories of the system
that  connect  all these three  Toda lattices.
This is a non-trivial illustration
of the fact  that the Hamiltonian
vector field is incomplete for every  Toda lattice  with ``bad signs''.

We wish to find the trajectory of the reduced system associated
with  some initial value $(u_2,\ldots, u_6)$ at $t=0$ parametrizing a point
$(g,J)\in \M^{\rm DS}=\M^{\rm red}$.
As was explained in  Remark 3, for this it is enough to solve the linear
equation
with constant coefficients in (6.15b), which now reads
$$
{d^3 u_6(t)\over dt^3} -u_2{d u_6(t)\over dt} -u_3 u_6(t)=0.
\eqno(7.29)$$
For simplicity, we choose to consider only those trajectories for which
the  characteristic equation corresponding to (7.29),
$$
y^3 - u_2 y - u_3 =\det\left( y{\bf 1}_3 - J(u_2,u_3)\right)=0,
\eqno(7.30)$$
has roots of the form
$$
y_1=-2a,\quad
y_2=a+b\sqrt{-1} ,\quad
y_3=a-b\sqrt{-1},\qquad
a,b\in \Real\setminus\{0\}.
\eqno(7.31)$$
We note that  this excludes the initial values in $\M^{\rm red}_e$,
since  the Lax matrix of the Toda lattice
with good sign has distinct {\it real} eigenvalues.

The existence of a pair of complex conjugate roots of (7.30) may
be ensured, for instance, by choosing
$$
u_2 <0,
\qquad
u_3\neq 0.
\eqno(7.32)$$
In fact,  if (7.32) holds, then  introducing   $r$ and $\vartheta$ by
$$
r:=-{\rm sign}(u_3)\sqrt{\left\vert{u_2\over 3}\right\vert},\qquad
\sinh\vartheta :=-{1\over 2} {u_3\over r^3},
\eqno(7.33{\rm a})$$
one has
$$
a=r\sinh{\vartheta\over 3},
\qquad
b=\sqrt{3} r\cosh{\vartheta\over 3}.
\eqno(7.33{\rm b})$$

The solution of (7.29) corresponding to roots of the form in (7.31)
can be written as
$$
u_6(t)=A e^{-2at} + e^{at}\left(B\sin{bt} + C \cos{bt}\right),
\eqno(7.34)$$
where $A$, $B$, $C$ are  real constants  determined by the
initial condition.
We shall not  need the explicit form of these constants,
only that for a generic initial condition they do not vanish,
which is obvious.
According to (6.15a), the complete solution is then given by
$$
u_2(t)=u_2,\quad
u_3(t)=u_3,\quad
u_4(t)={d^2 u_6(t)\over dt^2},\qquad
u_5(t)={d u_6(t)\over dt}\,.
\eqno(7.35)$$

We are interested in  the qualitative behaviour of the principal minors
$Q_3(u(t))$, $Q_2(u(t))$ of $g(u(t))$ along the above trajectory.
We have $Q_3(u(t))=u_6(t)$ and from  (7.34), (7.35) we can determine
$Q_2(u(t))$ as
$$
Q_2(u(t))=u_6(t) u_4(t)- u^2_5(t)=
\hat A e^{2at} +  e^{-at}\left( \hat B \sin{bt} +\hat C \cos{bt}\right).
\eqno(7.36)$$
The  computation yields
$$
\hat A=-2b^2 (B^2 + C^2).
\eqno(7.37)$$
The explicit form of  $\hat B$, $\hat C$ will not be needed.

It is important that the coefficient $\hat A$ of $e^{2at}$ in (7.36)
is {\it negative}.
Now we argue that
the coefficient $A$ of $e^{-2at}$ in (7.34) must also
be {\it negative}.
Indeed, if $A$ was positive, then $Q_3(u(t))$ would have the limit
$+\infty$ as $t$ tends to $-\left({\rm sign}( a)\right)\infty$.
On the other hand, since $\hat A$ is negative,
 $Q_2(u(t))$ oscillates around zero as $t$ tends to
$-\left({\rm sign} (a)\right)\infty$.
But then the trajectory would necessarily enter $\M^{\rm red}_e$,
where both $Q_2$ and $Q_3$ are positive.
This is impossible since
the initial value of the trajectory (at $t=0$)  is outside $\M_e^{\rm red}$,
which is disconnected from its complement.
This means that $A<0$ must necessarily hold for the generic
solution belonging to  characteristic eigenvalues of the form in (7.31).
(Actually, without using (7.37),
an extension of this  argument would in itself imply that both
$A$ and $\hat A$ must be negative.)

We have seen that the trajectory associated with
a generic initial  condition for which $J(u_2,u_3)$
has eigenvalues of the form in (7.31)  satisfies (7.34), (7.36)
with $A<0$, $\hat A<0$ and non-zero $B$, $C$, $\hat B$, $\hat C$.
This implies the following alternatives
for the asymptotic behaviour of such a trajectory:

\item{a)} {\it  If $a<0$, then
$$
\lim_{t\to +\infty}{Q_3(u(t))}=-\infty,
\qquad
\lim_{t\to -\infty}{Q_2(u(t))}=-\infty,
\eqno(7.38{\rm a})$$
and  $Q_3(u(t))$ (resp.~$Q_2(u(t))$) oscillates around $0$ for
large negative (resp.~positive) time.}

\item{b)} {\it  If $a>0$, then
$$
\lim_{t\to -\infty}{Q_3(u(t))}=-\infty,
\qquad
\lim_{t\to +\infty}{Q_2(u(t))}=-\infty,
\eqno(7.38{\rm b})$$
and  $Q_3(u(t))$ (resp.~$Q_2(u(t))$) oscillates around $0$ for
large positive (resp.~negative) time.}

\noindent
In either case, the trajectory oscillates
between a pair of connected components of the big cell
of $SL(3,\Real)$ as $\vert t\vert$ tends to $\infty$.
These pairs of  connected components consist of  the determinant 
one matrices
for which one of the principal minors is negative.
The pair in question is different for $t$ approaching
plus or minus infinity.
It follows  that the trajectory enters
all  the three  open submanifolds
$\M_m^{\rm red}\subset \M^{\rm red}$ for $m\neq e$
induced by the Bruhat decomposition.
 In particular, this confirms that the
Hamiltonian vector field of the Toda lattice
$(\M^{\rm red}_m,\Omega^{\rm red}_m, \H^{\rm red}_m)$
is incomplete for $m\neq e$.

Recall that  $J$ appearing in  $(g,J)\in \M^{\rm DS}$
is conjugate to the Lax matrix of a Toda lattice
if $g$ belongs to the big cell.
In conclusion, the qualitative behaviour found  above
is consistent with the general result that
the trajectory of the Toda lattice blows up
if the Lax matrix comprising the initial data admits a
complex eigenvalue \q{\GS,\KY}.

\bigskip
\medskip
\centerline{\bf 8.~Conclusion}
\medskip

In this paper we investigated  the natural  regularization of
incomplete Toda lattices associated with  $sl(n,\Real)$
that results from  Hamiltonian reduction.
We obtained a reduced phase space from  $T^*SL(n,\Real)$
which  contains an open dense
submanifold consisting of $2^{n-1}$ Toda lattices
and a complementary part consisting of lower
dimensional submanifolds serving to glue together
the blowing up trajectories of the Toda lattices.
We developed tools,  such as the double DS gauge
and the  hypersurface model of $\M^{\rm red}$,
 for further investigating the
global structure of the reduced system,
and used them to analyze the simplest
examples corresponding to $SL(n,\Real)$ for $n=2,3,4$
(see also Appendix A).
The results presented in the main text
for the Lie algebra $sl(n,\Real)$ can be generalized to an arbitrary
simple Lie algebra as explained  in Appendix B.

Much work remains to be done to explore
the structure of the  reduced system,
which appears quite interesting, and complicated.
For example, the Toda lattice with good sign, whose flow  is complete,
should occupy  a connected component of the phase space
whenever it is contained as a subsystem in the reduced system,
although this conjecture has been proven for $n=2,3,4$ only.
It  seems clear, but has not been shown yet,
that the  reduced phase space is not
the  cotangent bundle of  some configuration space in general.
Concerning the hypersurface model of $\M^{\rm red}$,
the regularity conjectures given in (6.18) and (6.25) should be verified
for arbitrary $n$.

A particularly  challenging problem is
to construct  the quantum mechanical version
of the reduced system and to determine, e.g.,  the joint
spectra of the operators
corresponding to the Casimirs of $sl(n,\Real)$.
The  quantization of  the
open Toda lattice  with good sign  is surveyed  in \q{\STS}
using a reduction method based on the Iwasawa decomposition.
One might try to extend this ``first quantize then reduce''
method to our case based on the Bruhat decomposition, or one might
try  to directly quantize the reduced classical system.
Direct quantization  has been attempted in \q{\F,\KT} for the simplest
case of $SL(2,\Real)$.

It could be also worthwhile to search for
a classification of singular solutions in the field theoretic 
version of the
open Toda lattices with the aid of the analogue of the Hamiltonian
reduction used here, which is described in \q{\FORTW}.
By direct methods, such classification was investigated in \q{\PP}
for  the Liouville equation related to $SL(2,\Real)$.

Finally, let us remind  that singular solutions
of many integrable systems  arise from the breakdown of solvability
in the factorization problem inherent in  the AKS scheme, and that
Hamiltonian reduction can provide a
regularization of such singularities in many instances 
(see \q{\first,\R,\second}).
In the present study we demonstrated
that  the reduced Hamiltonian systems
realizing  the regularization of singular Toda lattices
possess a  rich and intriguing structure.
We hope that our paper may
serve as a non-trivial illustration of
the aforementioned general   properties of integrable systems.


\def\A{{\rm A}}
\def\L{{\cal L}}
\def\T{{\cal T}}
\def\U{{\cal U}}
\def\deg{\hbox{deg}\,}

\bigskip\bigskip\bigskip
\centerline{\bf Appendix A: $\M^{\rm red}_{\pm e}$ are
connected components of  $\M^{\rm red}(I_-^+, I_+^+)$
for $SL(4,\Real)$}
\medskip

In Section 7 we discussed in some detail the
reduced systems for the two simplest examples  $SL(2,\Real)$ and
$SL(3,\Real)$.
If we go one step further,
i.e., to $SL(4,\Real)$, we learn from Proposition 1
that two non-isomorphic
reduced systems
$(\M^{\rm red},\Omega^{\rm red}, \H^{\rm red})_\pm$
are possible depending on which orbit of $\Delta$
the pair $(I_-, I_+)$ belongs to.  Proposition 2 then says that
the Toda lattice
with good sign (i.e., all $\mu_i>0$ in (4.1)) appears
only in the reduced system
$(\M^{\rm red},\Omega^{\rm red}, \H^{\rm red})_+$, and that it
appears in two copies among the eight Toda lattices allowed.
The aim of this appendix is to show that these
Toda lattices with good sign, which arise from the components of
the big cell of $SL(4,\Real)$ containing $\pm e$,
are  connected components of the reduced phase space.
This will provide a proof of the conjecture for $n = 4$
mentioned at the beginning of Section 7.
Our strategy will be similar to the one used for $SL(3, \Real )$.

We first recall that the double DS gauge introduced in Section 6
allows for parametrizing the general element
$(g,J)\in \M^{\rm red}(I_-^+,I_+^+)$ as
$$
J=\pmatrix{0&u_2&u_3&u_4\cr1&0&0&0\cr0&1&0&0\cr0&0&1&0\cr}
\qquad\hbox{and}\qquad
g=\pmatrix{g_{1,1}&g_{1,2}&g_{1,3}&u_5\cr
            g_{1,2}&g_{1,3}&u_5&u_6\cr
            g_{1,3}&u_5&u_6&u_7\cr
            u_5&u_6&u_7&u_8\cr
},
\eqno(\A.1{\rm a})
$$
where
$$
\eqalign{
g_{1,1} & = u_2 (u_2 u_6 + u_3 u_7 + u_4 u_8)  + u_3 u_5 + u_4 u_6, \cr
g_{1,2} & = u_2 u_5 + u_3 u_6 + u_4 u_7, \cr
g_{1,3} & = u_2 u_6 + u_3 u_7 + u_4 u_8.
}
\eqno(\A.1{\rm b})
$$
These parameters $u = (u_2, \ldots, u_8)$ are subject to
the condition
$P(u) = \det g(u)=1$.
The set of globally defined functions in (5.15)
$$
Q_4(g,J) = u_8, \quad
Q_3(g,J)=\det\pmatrix{u_6&u_7\cr u_7&u_8\cr}, \quad
Q_2(g,J)=\det\pmatrix{g_{1,3}&u_5&u_6\cr u_5&u_6&u_7\cr u_6&u_7&u_8\cr}\,,
\eqno(\A.2)
$$
then give the mapping
$Q:=(Q_2,Q_3,Q_4): \M^{\rm red}\rightarrow \Real \times \Real \times \Real$.
Using this we can write the open submanifolds
$\M^{\rm red}_{\pm e}\subset \M^{\rm red}$,
each of which carries the Toda lattice with good sign, as
$$
\M^{\rm red}_{\pm e} = Q^{-1}\left(\Real^\pm \times \Real^\pm
\times \Real^\pm \right).
\eqno(\A.3)
$$

To see that one of the submanifolds, say $\M^{\rm red}_{e}$,
is topologically disconnected from its complement
$\M^{\rm red}\setminus \M^{\rm red}_e$, let us first assume that
this is not true.  Then
we can consider a continuous path $\gamma(t) =
(g(t), J(t)) \in \M^{\rm red}$ for $t\in  [t_0, t_0+\epsilon ]$
($t_0\in \Real$ and $\epsilon >0$)
connecting one point from $\M^{\rm red}_{e}$ and another from
$\left(\M^{\rm red}\setminus\M^{\rm red}_e\right)\cap\M^{\rm red}_{\rm low}$
in such a way that
$$
Q_2(\gamma(t))>0,\quad Q_3(\gamma(t))>0, \quad Q_4(\gamma(t))>0,
  \qquad \hbox{for}\quad t>t_0
\eqno(\A.4)
$$
and
$$
Q_2(\gamma(t_0))Q_3(\gamma(t_0))Q_4(\gamma(t_0))=0.
\eqno(\A.5)
$$
In what follows we shall prove that in fact there exists no
path $\gamma(t)$  satisfying (A.4) and (A.5),
which implies that our assumption is wrong and hence
$\M^{\rm red}_{e}$ must be disconnected from its complement.
Our proof consists of two parts.  In the first part,
we show that (A.4) and (\A.5) actually imply
$$
Q_2(\gamma(t_0))=Q_3(\gamma(t_0))=Q_4(\gamma(t_0))=0.
\eqno(\A.6)
$$
In the second part we shall show that (A.4) and (A.6) lead to a
contradiction.

We begin the first part of the proof by noting that
(A.4) and (A.5) imply either
$$
Q_4(\gamma(t_0))=0,
\qquad
Q_3(\gamma(t_0))\geq 0,
\qquad
Q_2(\gamma(t_0))\geq 0,
\eqno(\A.7\hbox{a})
$$
or
$$
Q_4(\gamma(t_0))\geq 0,
\qquad
Q_3(\gamma(t_0)) = 0,
\qquad
Q_2(\gamma(t_0))\geq 0,
\eqno(\A.7\hbox{b})
$$
or the case where $Q_2$ and $Q_4$ are interchanged in (A.7a).
However, it is
enough to consider only the two cases (A.7a), (A.7b)
since one can convert the last case into (A.7a)
using the involutive symmetry transformation presented
in Section 5 (see Proposition 4).

Now consider the case (A.7a).  Since
$$
0 = Q_4(\gamma(t_0)) = u_8(t_0),
\eqno(\A.8)
$$
we have
$Q_3(\gamma(t_0)) = - u_7^2(t_0) \leq 0$.
{}From (A.7a) we find
$u_7(t_0) = 0$ and
$$
Q_3(\gamma(t_0)) = 0.
\eqno(\A.9)
$$
It then follows that
$Q_2(\gamma(t_0)) = - u_6^3(t_0)$ and therefore from (A.7a) that
$$
u_6(t_0) \leq 0.
\eqno(\A.10)
$$
On the other hand, at $t > t_0$, condition (A.4) requires
$$
0 < Q_4(\gamma(t)) = u_8(t)
\quad \hbox{and} \quad
0 < Q_3(\gamma(t)) = u_6(t) u_8(t) - u_7^2(t)\,,
\eqno(\A.11)
$$
and hence
$$
u_6(t) > 0 .
\eqno(\A.12)
$$
Due to the continuity of the path
this is consistent with (A.10)
only if $u_6(t_0) = 0$, i.e.,
$$
Q_2(\gamma(t_0)) = 0
\eqno(\A.13)
$$
as well.  We have thus learned that (A.7a) reduces to (A.6).

Next, consider the case (A.7b) for which
$$
0 = Q_3(\gamma(t_0))
= u_6(t_0) u_8(t_0) - u_7^2(t_0).
\eqno(\A.14)
$$
{}From this we find
$$
\eqalignno{
Q_4^3(\gamma(t_0)) Q_2(\gamma(t_0)) &=
    u_8^3(t_0) \, \det \pmatrix{ g_{1,3}   &   u_5 &  u_6 \cr
              u_5 &   u_6 &  u_7 \cr
              u_6 &   u_7 &  u_8 \cr}(t_0) \cr
    &= - \bigl\{ u_7^3(t_0) - u_5(t_0) u_8^2(t_0) \bigr\}^2 \leq 0.
&(\A.15)
}
$$
On account of (A.7b) the equality must hold in (A.15) and,
accordingly, either $Q_4(\gamma(t_0))$ or $Q_2(\gamma(t_0))$ must be zero.
This allows us to go back to the previous case (A.7a)
(or the last case
which is equivalent to (A.7a) due to the
involutive symmetry),
and thus we conclude that (A.7b) also reduces to (A.6).
This completes the first part of our proof.

We here remind  that on a smooth manifold every continuous path
between two fixed points
is continuously deformable to a smooth (that is $C^\infty$) path.
In particular, if there is a continuous
path on $\M^{\rm red}$   satisfying (A.4) and (A.5),
then there also exists a smooth path meeting these conditions.

For the second part of the  proof we need to introduce some
notations.  First, given a $C^\infty$ function $f(t)$
of $t \in [t_0, t_0 + \epsilon]$, let
$\L(f)$ be the leading term of the function $f$ in the
formal Taylor expansion at $t = t_0$.
That is, if ${{d^i f}\over {d t^i}}(t_0) \neq 0$ for some $i\geq 0$,
$$
\L(f) := {1\over{n!}}{{d^n f}\over {d t^n}}(t_0)\,(t-t_0)^n,
\eqno(\A.16{\rm a})
$$
where
$$
n = \deg(f) := \hbox{min}\,\Bigl\{\,i\,\Big\vert\,
{{d^i f}\over {d t^i}}(t_0) \neq 0 \Bigr\}.
\eqno(\A.16{\rm b})
$$
We put $\deg(f)= \infty$ and $\L(f) = 0$ if
${{d^i f}\over {d t^i}}(t_0) = 0$ for all $i$.
It is obvious that
$$
\deg(fg)=\deg(f)+\deg(g), \qquad \L(fg) = \L(f)\L(g),
\eqno(\A.17)
$$
for arbitrary smooth functions $f$, $g$.
If $f$ is written as a sum of smooth functions,
$f = \sum_{i = 1}^m f_i$, then we define
$$
\T(f\vert\, f_1, \ldots, f_m) :=
\sum_{i = 1}^m \delta_{\sigma(i),k} \L(f_i)
\eqno(\A.18{\rm a})
$$
where
$$
\sigma(i) := \deg(f_i) \qquad \hbox{and}
\qquad
k := \hbox{min}\,\{ \deg(f_1), \ldots, \deg(f_m) \}.
\eqno(\A.18{\rm b})
$$
Note that
$$
a\, \T(f\vert\, f_1, \ldots, f_m) = \T(af\vert\, a f_1, \ldots, a f_m),
\qquad\forall\, a\in \Real.
\eqno(\A.19)
$$
It should be stressed that $\T(f\vert\, f_1, \ldots, f_m)$ is defined with
respect to the set of functions $\{f_1, \ldots, f_m\}$,
and in general the outcome of the operation depends on
the set\footnote{${}^{5}$}
{\ninerm For example, for $f = f_1 + f_2$ with
$f_1 = (t-t_0) + (t-t_0)^2$ and
$f_2 = - (t-t_0)$, we have
$\T(f\vert\, f)=\L(f) = (t-t_0)^2$ whereas
$\T(f\vert\, f_1, f_2) = \L(f_1) + \L(f_2)=0$.
}.
Upon specifying the set, we may
write $\T(f_1+\cdots +f_m)$ or simply
$\T(f)$ for $\T(f\vert\, f_1, \ldots, f_m)$.
{}From the definitions, it is easy to show that
$$
f(t) > 0 \quad\hbox{for all $t\in (t_0, t_0+\epsilon]$}
\quad \Rightarrow \quad \L(f) \geq 0
\quad \hbox{and} \quad
\T(f\vert\, f_1,\ldots, f_m) \geq 0.
\eqno(\A.20)
$$
Here $\L(f)=0$ if and only if $\deg(f)=\infty$.
Since the operation $\T(f\vert\, f_1, \ldots, f_m)$ in (A.18)
keeps only the leading
term(s) among the $\L(f_i)$ for  $i = 1, \ldots, m$,
$\T(f)$ vanishes if the non-zero leading terms cancel each other.

Let us now return to our reduced system.  Recall that (A.6) implies
$u_8(t_0) = u_7(t_0) = u_6(t_0) = 0$, and hence $u_5(t_0) = \pm 1$
from the condition $P(u(t_0)) = \det g(t_0) = 1$.
On account of the fact that
$$
dP(u)\vert_{P(u) = 1} = \pm(4 du_5 - 2 u_2 du_7 - u_3 du_8) \ne 0
\qquad\hbox{at}\quad u = u(t_0),
\eqno(\A.21)
$$
we see that the set $\{u_2, u_3, u_4, u_6, u_7, u_8\}$
can be taken as  coordinates on a neighbourhood
$\U$ of the point $\gamma(t_0)$ diffeomorphic to an open ball in $\Real^6$.
In $\U\subset \M^{\rm red}$, we can consider
the {\it generic} class of smooth paths satisfying (A.4), (A.5)
as well as the  following additional condition
for the coordinate functions along the path,
$$
\L(u_i)\neq 0 \qquad\hbox{for}\qquad  i=2,3,4, 6, 7,8.
\eqno(\A.22)$$
This  class of paths is {\it not} empty,
since, as is intuitively clear,  every smooth path satisfying (A.4), (A.5)
can be deformed to a smooth path connecting the same end points
in such a way that (A.22) holds in addition to (A.4), (A.5).
The class of paths satisfying (A.22) is generic
among the smooth paths satisfying (A.4) and (A.5), because
(A.22) is stable under small deformations.

Let now $\gamma(t)$  be an arbitrary  smooth path
subject to the requirements (A.4), (A.5), (A.22).
If we write $u_5(t) = \pm 1 + \hat u_5(t)$, then
the functions $u_8(t), u_7(t), u_6(t), \hat u_5(t), g_{1,3}(t)$
all vanish at $t = t_0$, and
hence the degree defined in (A.16b) is non-zero for each of these
functions.
In addition, the degrees of $u_6(t)$, $u_7(t)$, $u_8(t)$ are finite
according to (A.22).
In terms of these functions the first condition in (A.4) reads
$$\eqalign{
0 < Q_2(\gamma(t)) = -& u_8 \pm 2u_6u_7 - u_6^3 \cr
&+  g_{1,3}u_6u_8 -
g_{1,3}u_7^2 \mp 2\hat u_5 u_8 +2\hat u_5 u_6u_7 - \hat u_5^2 u_8.\cr}
\eqno(\A.23)
$$
For brevity we hereafter suppress
the parameter $t$ in the functions.
Applying the operations (A.16), (A.18) on the two inequalities in (A.11),
respectively, we find
$$
\L(u_8) > 0,
\eqno(\A.24)
$$
and
$$ \T(Q_3\vert\,u_6u_8-u_7^2)=\T(u_6u_8 - u_7^2) \geq 0,
\eqno(\A.25)
$$
where we used (A.20) and (A.22).
{}From (A.25) we  now notice that
$$
\deg(u_7^2) \geq \deg(u_6u_8),
\eqno(\A.26)
$$
since otherwise $\T(u_6u_8 - u_7^2) = - \{\L(u_7)\}^2 < 0$.
On the other hand, we see that,
since $\deg(g_{1,3}) \geq 1$,
the degree of any of the five terms
in the second line of (A.23) is higher
than the degree of at least
one of the first three terms $\{ -u_8, \pm 2u_6u_7, - u_6^3\}$.
Indeed, the degrees of those
three terms in the second line that have the factor $u_8$
are obviously higher than that of $u_8$, while
$\deg(g_{1,3} u_7^2) \geq \deg(g_{1,3} u_6u_8) > \deg(u_8)$ and
$\deg(\hat u_5 u_6u_7) > \deg(u_6u_7)$.
Thus, applying (A.18) on (A.23)
and using (A.20) we obtain that
$$
\T(Q_2)=\T(-u_8 \pm 2u_6u_7 - u_6^3) \geq 0.
\eqno(\A.27)
$$

To be more explicit, let us put
$$
\L(u_6) = a(t-t_0)^m, \qquad
\L(u_7) = b(t-t_0)^n, \qquad
\L(u_8) = c(t-t_0)^l,
\eqno(\A.28)
$$
where $a$, $b$, $c$ are non-vanishing  constants and
$m$, $n$, $l$ are some positive integers.
Observe then that (A.12) and (A.22) imply
$\L(u_6) > 0$, that is
$$
a > 0,
\eqno(\A.29)
$$
and therefore from (A.19) and (A.27) we have
$$
0 \leq a\,\T(-u_8 \pm 2u_6u_7 - u_6^3)
         = - \T\bigl(ac (t-t_0)^{l} \mp 2 a^2 b (t-t_0)^{m+n}
                    + a^4 (t-t_0)^{3m}\bigr).
\eqno(\A.30)
$$
Observe also that (A.24) is equivalent to
$$
c > 0,
\eqno(\A.31)
$$
and that (A.26) implies
$$
m+l \leq 2n.
\eqno(\A.32)
$$

Now suppose $m + n < 3m$.  From (A.32) we then have $l < m + n $
and hence
$$
a\,\T(-u_8 \pm 2u_6u_7 - u_6^3) = - ac (t-t_0)^{l} < 0,
\eqno(\A.33)
$$
which contradicts with (A.30).
Thus we find
$$
m + n \geq 3m.
\eqno(\A.34)
$$
However, $m + n > 3m$ is impossible since it leads
to either (A.33) for $l < 3m$, or
$$
a\,\T(-u_8 \pm 2u_6u_7 - u_6^3) =  - a^4 (t-t_0)^{3m} < 0,
\eqno(\A.35)
$$
for $l > 3m$, or
$$
a\,\T(-u_8 \pm 2u_6u_7 - u_6^3) = - (ac + a^4) (t-t_0)^{3m} < 0,
\eqno(\A.36)
$$
for $l = 3m$, all of which contradict with (A.30).
Thus we must have $m + n = 3m$, and combining this with
(A.32) we get $l \leq 3m$.  But since $l < 3m$
leads again to (A.33) we conclude
that the three terms in (A.30)
are of the same degree, $l = m + n = 3m$, i.e.,
$$
l = 3m \qquad \hbox{and} \qquad n = 2m.
\eqno(\A.37)
$$

Having found the ratio of the three degrees,
we see that (A.25) reads
$$
\T(u_6u_8 - u_7^2)=
(ac - b^2)(t-t_0)^{4m} \geq 0.
\eqno(\A.38)
$$
It then follows that
$$
a\,\T(-u_8 \pm 2u_6u_7 - u_6^3)
  =  - \bigl\{(ac - b^2) + (b \mp a^2)^2 \bigr\}(t-t_0)^{3m} \leq 0.
\eqno(\A.39)
$$
Comparing this with (A.30) we  find that the equality must hold,
and accordingly
$$
\T(-u_8 \pm 2u_6u_7 - u_6^3) = 0.
\eqno(\A.39)
$$
Thus the leading terms of the first three terms in (A.23)
cancel each other, leaving higher degree terms.
Therefore the degree of the function $Q_2$ in (A.23)
necessarily satisfies
$$
\deg(Q_2) > 3m.
\eqno(\A.41)
$$
On the other hand, we have
$$
\deg(Q_4) = \deg(u_8) = 3m.
\eqno(\A.42)
$$

In the above we have proven that
$$
\deg(Q_2)>\deg(Q_4)
\eqno(\A.43)$$
for any smooth path $\gamma(t)$  satisfying (A.4), (A.5) and (A.22).
However, this is in conflict with the involutive symmetry
$\hat \psi_{(D_+,D_+)}:\M^{\rm red}\rightarrow \M^{\rm red}$
described in Section 5, which shows that if $\gamma(t)$
is a path connecting $\M^{\rm red}_e$
to its complement then there exists also another path
$\hat \gamma(t)$,
given by $\hat \gamma(t)=\hat \psi_{(D_+,D_+)}(\gamma(t))$,
that does the same in such a way that
$Q_i(\hat \gamma(t))=Q_{4+i-2}(\gamma(t))$.
Applying this symmetry to a smooth path
satisfying (A.4), (A.5), (A.22) and $\deg(Q_2)>\deg(Q_4)$
generically results in a smooth path satisfying
(A.4), (A.5), (A.22) and $\deg(Q_2) < \deg(Q_4)$.
This implies  that there must also exist a smooth path
satisfying (A.4), (A.5), (A.22) and $\deg(Q_2)<\deg(Q_4)$.
But this is impossible since we proved that (A.4), (A.5) and (A.22)
imply (A.43).
We conclude from this contradiction  that $\M^{\rm red}_e$ is indeed
disconnected from its complement.

Having shown that $\M^{\rm red}_{e}$ is a connected
component of $\M^{\rm red}$, the symmetry given in (5.14)
shows that $\M^{\rm red}_{-e}$ is also a connected  component.

We expect  that the Toda lattice with good sign is realized as a
connected  component whenever  it is contained in the reduced phase space
associated with   $SL(n,\Real)$, for  any $n$.
However, a new idea is required for proving this conjecture, since the
method used above would become impractical  for $n>4$.

\def\b{{\rm B}}

\bigskip
\medskip
\centerline{\bf Appendix B:
Generalization for  arbitrary  simple Lie algebras}
\medskip

The framework for regularizing Toda lattices
with ``bad signs'' described  in the
main text  for $sl(n,\Real)$ can be straightforwardly generalized for
an arbitrary  simple Lie algebra.
We now briefly present  this  generalization.

Let $\G$ be the normal (split) real form of some complex 
simple Lie algebra.
Then $\G$ is generated by the  Chevalley generators
$$
h_{\alpha_i}, \qquad e_{\alpha_i}, \qquad e_{-\alpha_i}
\eqno(\b.1)$$
associated with  the
simple roots $\alpha_i$ for $i=1,\ldots, r:={\rm rank}(\G)$.
The Chevalley involution $\theta$ of $\G$ operates as
$$
\theta(h_{\alpha_i})=-h_{\alpha_i},
\qquad
\theta(e_{\alpha_i})=-e_{-\alpha_i}.
\eqno(\b.2)$$
We have
${\rm tr}\left(e_{\alpha_i}e_{-\alpha_i}\right)>0$
with ``${\rm tr}$'' being  the  Killing form of $\G$ up to a 
positive constant.
In the triangular decomposition of equation (2.1) now
$\G_0={\rm span}\{ h_{\alpha_i}\}_{i=1}^r$ is the splitting  Cartan
subalgebra of $\G$ and $\G_{>0}$ (resp.~$\G_{<0}$) are the subalgebras
spanned  by the positive (resp.~negative) root vectors.

The phase space of the Toda lattice
of our interest is
$M_e=\G_0\times \G_0\simeq \O$ given as
in (2.18), now using
$$
I_+=\sum_{i=1}^{r} \nu_i^+ e_{\alpha_i},
\qquad
I_-=\sum_{i=1}^{r} \nu_i^- e_{-\alpha_i},
\qquad
\nu_i^\pm \neq 0.
\eqno(\b.3)$$
It has the  symplectic form  $\omega_e=d{\rm tr}\left(pdq\right)$ and,
with $\nu_i := \nu_i^- \nu_i^+\tr \left( e_{\alpha_i}e_{-\alpha_i}\right)$,
the Hamiltonian
$$
H_e(q,p):={1\over 2}\tr\left(p^2\right)+
\tr\left( I_- e^q I_+ e^{-q}\right)
={1\over 2}\tr\left(p^2\right)+\sum_{i=1}^r\nu_i e^{\alpha_i(q)}.
\eqno(\b.4)$$
The Toda lattice $(M_e, \omega_e, H_e)$, which is 
singular if
$\nu_i<0$ for some $i$,
is contained in the reduced system following from a Hamiltonian reduction
of the natural system $(\M,\Omega,\H)$ on $\M=T^*G$
with $G$ being a connected Lie group corresponding to $\G$.
To define the reduction, $N_+$ and $N_-$ in (3.8)
are now taken to be   the Lie subgroups of $G$ associated with
$\G_{>0}$ and $\G_{<0}$, respectively,
the constrained manifold $\M^{\rm c}$ is defined similarly to (3.11),
and $\M^{\rm red}=\M^{\rm c}/N$.
We wish  to sketch the structure of
$(\M^{\rm red}, \Omega^{\rm red}, \H^{\rm red})$.

Let $K$ be the Lie subgroup of $G$ corresponding to
the Lie subalgebra of $\G$ given by the fixed point set of  $\theta$.
Let ${\bf M}^*\subset K$ be the {\it normalizer} and
${\bf M}\subset K$ the {\it centralizer} of $\G_0$ with respect to the
adjoint representation of $G$ on $\G$ restricted to $K\subset G$.
Then ${\bf W}:={\bf M}^*/{\bf M}$ is the Weyl group of $\G$ with
respect to the
Cartan subalgebra $\G_0$.
Denote by $A$ the Lie subgroup of $G$ corresponding to $\G_0$.
We have (see \q{\He,\War})
the {\it Bruhat decomposition} of $G$:
$$
G=\cup_{w\in {\bf W}} N_+ {\bf M} m^*_w A N_- \qquad \hbox{(disjoint union)},
\eqno(\b.5)$$
where $m^*_w\in {\bf M}^*$ is an arbitrary representative of $w\in {\bf W}$.
The big cell belongs to the identity element of ${\bf W}$.
It is an open, dense submanifold in $G$ having the
 connected components
$G_m:= N_+ m A N_-$ for all  $m\in {\bf M}$,
which are diffeomorphic to $N_+ \times A\times N_-$.
The decomposition in  (\b.5) can be rewritten as
$$
G=\cup_{m\in {\bf M}} G_m \cup G_{\rm low},
\qquad
G_{\rm low}=\cup_{w\neq e} N_+ {\bf M} m^*_w A N_-.
\eqno(\b.6)$$
The lower dimensional submanifold  $N_+ {\bf M} m^*_w A N_-\subset G$ is
diffeomorphic to
the product of the factors on the left and right hand sides
of the equality
$$
N_+^w {\bf M} m^*_w A N_-=N_+ {\bf M} m^*_w A N_-=N_+ {\bf M} m_w^* AN_-^w
\eqno(\b.7{\rm a})$$
with
$$
N_+^w = N_+\cap m_w^* N_+ (m_w^*)^{-1}
\quad\hbox{and}\quad
N_-^w = N_-\cap (m_w^*)^{-1} N_- m_w^*.
\eqno(\b.7{\rm b})$$

In the same way as we saw  in Section 3, the
Bruhat decomposition of $G$ induces a
decomposition of $\M^{\rm red}$ of the form in (3.17).
For $m=e$ the subsystem 
$(\M_m^{\rm red}, \Omega_m^{\rm red}, \H^{\rm red}_m)$
is isomorphic to $(M_e, \Omega_e, H_e)$,
otherwise it is a Toda lattice of the same kind obtained by simply
replacing $I_+$ with $I_+^m=m I_+ m^{-1}$.
In particular, the Hamiltonian $\H^{\rm red}_m$ of this Toda lattice
has the form
$$
\H_m^{\rm red}(q,p):={1\over 2}\tr\left(p^2\right)+
\tr\left( I_- e^q I^m_+ e^{-q}\right)
={1\over 2}\tr\left(p^2\right)+\sum_{i=1}^r\mu_i e^{\alpha_i(q)},
\quad \mu_i=m_{\alpha_i} \nu_i,
\eqno(\b.8)$$
where $m_{\alpha_i}$ is defined by
$me_{\alpha_i}m^{-1}=m_{\alpha_i} e_{\alpha_i}$.
Therefore $\M^{\rm red}$ contains Toda lattices  that are in
general singular (incomplete)
but are glued together in such a way that
their  singularities are   regularized in the entire system.
With the aid of equations  (6.1) and (6.2),
one can  introduce the double DS gauge $\M^{\rm DS}\subset \M^{\rm c}$
to provide a globally valid model of  $\M^{\rm red}$,
which could  be useful for exploring  the  topology of $\M^{\rm red}$.

\medskip
\noindent
{\it Remark 6.}
The analogue of the splitting in (2.22) is given by
$\G={\cal K}+ \G_{\leq 0}$, where ${\cal K}$ is
the Lie algebra of $K\subset G$.
Thus one has an orbital model $\tilde \O$ of the Toda
lattice  with  good sign  similar to (2.27),
where one must now take $I_-=-\theta(I_+)$.
This Toda lattice arises as the reduced system  from a Hamiltonian
reduction of
the system $(T^*G, \Omega, \H)$ based on  the symmetry group
$K\times N_-$, whose action on $T^*G$ is induced by the left
translations of $K$ on $G$ and the right translations of $N_-$
on $G$.
The reduction is defined fixing the value of the momentum
map of the $K$--action to zero, and fixing the momentum map
of the $N_-$--action to the value $I:=(I_+ -\theta(I_-))$.
Using the identifications
${\cal K}^*= \left(\G_{\leq 0}\right)^\perp$ and
$\left(\G_{\leq 0}\right)^*={\cal K}^\perp$, here
$I\in {\cal K}^\perp$ represents a functional on
$\G_{<0}\subset \G_{\leq 0}$.
This treatment of the Toda lattice relates to
the globally valid Iwasawa decomposition, $G=KAN_-$,
of the group $G$.
It reproduces  the Hamiltonian reduction treatment
given in \q{\OP,\P,\FO} if one first performs the reduction
of the $K$--symmetry, which leads to geodesic motion
on the symmetric space $K\backslash G \simeq AN_-$.

\bigskip\bigskip\bigskip
\centerline{\bf Acknowledgments}
\medskip

L.F.~wishes  to thank  H.~Flaschka for
a discussion  and for drawing his attention
to some references.  He is also grateful for the
hospitality received while he was
a visitor at the Laboratoire de Physique Th\'eorique, ENS Lyon, and at
the Department of Physics, University of Wales Swansea,
where part of this work was done.

\vfill\eject

\baselineskip=15pt
\parskip 6pt
\centerline{\bf References}
\vskip 5pt

\bibitem{\BFP}
J.\  Balog, L.\ Feh\'er, L.\ Palla,  in preparation.

\bibitem{\DLNT}
P.\ Deift, L.C.\ Li, T.\ Nanda, C.\ Tomei,
{\it The Toda flow on a generic orbit is integrable},
Comm.\ Pure  Appl.\ Math.\ {\bf 39} (1986) 183-232.

\bibitem{\DS}
V.G.\ Drinfeld, V.V.\ Sokolov,
{\it Lie algebras and equations of KdV type},
J.\ Sov.\ Math.\ {\bf 30} (1985) 1975-2036.

\bibitem{\EFS}
N.M.\ Ercolani, H.\ Flaschka, S.\ Singer,
{\it The geometry of the full Kostant-Toda lattice},
pp.\ 181-225 in: Birkh\"auser Progr.\ in Math.\ Vol.\  115.

\bibitem{\F}
T.\ F\"ul\"op,
{\it Reduced $SL(2,\Real)$ WZNW quantum mechanics},
ITP Budapest Report 509, hep-th/9502145 (to appear in J.\ Math.\ Phys.).

\bibitem{\FH}
H.\ Flaschka, L.\ Haine,
{\it Vari\'et\'es de drapeaux et r\'eseaux de Toda},
Math.\ Z.\ {\bf 207} (1991) 545-556.

\bibitem{\FO}
L.A.\ Ferreira, D.I.\ Olive,
{\it Non-compact symmetric spaces and the Toda molecule equations},
Commun.\ Math.\ Phys.\ {\bf 99} (1985) 365-384.

\bibitem{\FORTW}
L.\ Feh\'er, L.\ O'Raifeartaigh, P.\ Ruelle, I.\ Tsutsui, A.\ Wipf,
{\it
On Hamiltonian reductions of the Wess-Zumino-Novikov-Witten theories},
Phys.\ Rep.\  {\bf 222} (1992) 1-64.

\bibitem{\GS}
M.I.\ Gekhtman, M.Z.\ Shapiro,
{\it Completeness of real Toda flows and totally positive matrices},
Weizmann Institute of Science preprint.

\bibitem{\He}
S.\  Helgason,  {\sl Differential Geometry,
Lie Groups and Symmetric Spaces}, Chapter IX,
Academic Press, New York, 1978.

\bibitem{\KT}
H.\ Kobayashi, I.\ Tsutsui,
{\it Quantum mechanical Liouville model with attractive potential},
INS  preprint, INS-Rep.-1124.

\bibitem{\KY}
Y.\ Kodama, J.\ Ye,
{\it Toda hierarchy with indefinite metric},
Ohio State University preprint,  solv-int/9505010.

\bibitem{\OP}
M.A.\ Olshanetsky, A.M.\ Perelomov,
{\it Toda lattice as a reduced system},
Theor.\ Math.\ Phys.\ {\bf 45} (1981) 843-854.

\bibitem{\P}
A.M.\ Perelomov, {\sl Integrable Systems of Classical Mechanics and
Lie Algebras}, Birkh\"auser, Basel, 1990.

\bibitem{\PP}
A.K.\ Pogrebkov, M.K.\ Polivanov,
{\sl Sov. Sci. Rev. C. (Math. Phys.)} Vol.~5. (1985) pp. 197-272,
and references therein.

\bibitem{\R}
A.G.\ Reyman,
{\it Integrable Hamiltonian systems connected with graded Lie algebras},
J.\ Sov.\ Math.\ {\bf 19} (1982) 1507-1545.

\bibitem{\first}
A.G.\ Reyman, M.A.\ Semenov-Tian-Shansky,
{\it Reduction of Hamiltonian systems, affine Lie algebras and Lax equations},
Inventiones math.\ {\bf 51} (1979) 81-100.

\bibitem{\second}
A.G.\ Reyman, M.A.\ Semenov-Tian-Shansky,
{\it Group-Theoretical Methods in the Theory of Finite-Dimensional
Integrable-Systems},  in: {\sl Encyclopedia of
Mathematical Sciences, Vol.\ 16},
V.I.\ Arnold, S.P.\ Novikov (Eds.), Springer-Verlag, New York, 1994.

\bibitem{\STS}
M.A.\ Semenov-Tian-Shansky,
{\it Quantization of Open Toda Lattices},  in: {\sl Encyclopedia of
Mathematical Sciences, Vol.\  16},
V.I.\ Arnold, S.P.\ Novikov (Eds.), Springer-Verlag, New York, 1994.

\bibitem{\TF}
I.\ Tsutsui, L.\ Feh\'er,
{\it Global aspects of the WZNW reduction to Toda theories},
Prog.\ Theor.\ Phys.\ Suppl.\ {\bf 118} (1995) 173-190.

\bibitem{\War}
G.\ Warner,
{\sl Harmonic Analysis on Semi-Simple Lie Groups I},
Section 1.2, Springer-Verlag, New York, 1972.

\bye